\documentclass[12pt,preprint,dvips,deluxetable]{aastex}

\def \msun{$\mathrm{M}_\odot$}
\def \rsun{$\mathrm{R}_\odot$}

\def \kms{km~s$^{-1}$}
\def \1pap{Paper I}
\def \pap2{Paper II}
\def \deg{$^\circ$}

\makeatletter
\newcommand\cutinheadb[1]{%
 \noalign{\vskip .8ex}%
 \@ptabularcr
 \noalign{\vskip 0.1ex}%
 \multicolumn{\pt@ncol}{c}{#1}%
 \@ptabularcr
 \noalign{\vskip 1ex}%
 \hline
 \@ptabularcr
 \noalign{\vskip -1.5ex}%
}%
\def\cutinheadb@ppt#1{%
 \noalign{\vskip .8ex}%
 \@ptabularcr
 \noalign{\vskip 0.11ex}
 \multicolumn{\pt@ncol}{c}{#1}%
 \@ptabularcr
 \noalign{\vskip 1ex}%
 \hline
 \@ptabularcr
 \noalign{\vskip -1.5ex}%
}%
\makeatother

\makeatletter
\newcommand\cutinheadc[1]{%
 \noalign{\vskip .8ex}%
 \@ptabularcr
 \noalign{\vskip -4ex}%
 \multicolumn{\pt@ncol}{c}{#1}%
 \@ptabularcr
 \noalign{\vskip .8ex}%
 \hline
 \@ptabularcr
 \noalign{\vskip -1.5ex}%
}%
\def\cutinheadc@ppt#1{%
 \noalign{\vskip .8ex}%
 \@ptabularcr
 \noalign{\vskip -1.5ex}
 \multicolumn{\pt@ncol}{c}{#1}%
 \@ptabularcr
 \noalign{\vskip .8ex}%
 \hline
 \@ptabularcr
 \noalign{\vskip -1.5ex}%
}%
\makeatother

\author{Daniel~C.~Kiminki\altaffilmark{1}\altaffilmark{2},
Henry~A.~Kobulnicky\altaffilmark{1}, Ian~Ewing\altaffilmark{1},
Megan~M.~Bagley~Kiminki\altaffilmark{2},
Michael~Lundquist\altaffilmark{1}, Michael~Alexander\altaffilmark{1},
Carlos~Vargas-Alvarez\altaffilmark{1},
Heather~Choi\altaffilmark{1}, \&\ C.~B.~Henderson\altaffilmark{3}}

\altaffiltext{1}{Dept. of Physics \& Astronomy, University of Wyoming,
Laramie, WY 82070}
\altaffiltext{2}{Dept. of Astronomy, University of Arizona, 
Tucson, AZ 85721}
\altaffiltext{3}{Dept. of Astronomy, Ohio State University, 
Columbus, OH 43210}
\begin{document}

\title{Additional Massive Binaries in the Cygnus OB2 Association}

\begin{abstract}
We report the discovery and orbital solutions for two new OB binaries
in the Cygnus~OB2 Association, MT311 (B2V~$+$~B3V) and MT605
(B0.5V~$+$~B2.5:V).  We also identify the system MT429 as a probable
triple system consisting of a tight eclipsing 2.97 day B3V+B6V pair
and a B0V at a projected separation of 138 AU.  We further provide the
first spectroscopic orbital solutions to the eclipsing, double-lined,
O-star binary MT696 (O9.5V~$+$~B1:V), the double-lined, early B binary
MT720 (B0-1V~$+$~B1-2V), and the double-lined, O-star binary MT771
(O7V~$+$~O9V).  These systems exhibit orbital periods between 1.5 days
and 12.3 days, with the majority having P$<$6 days.  The two new
binary discoveries and six spectroscopic solutions bring the total
number of known massive binaries in the central region of the
Cygnus~OB2 Association to 20, with all but two having full orbital
solutions.
\end{abstract}

\keywords{techniques: radial velocities --- binaries: general ---
 (stars:) binaries: spectroscopic --- (stars:) binaries:
 (\textit{including multiple}) close --- stars: early-type --- stars:
 kinematics --- surveys }

\section{Introduction}
The ubiquity of close massive binaries in open clusters is becoming
well documented (e.g., \citealt{Sana09}, \citealt{Mahy09},
\citealt{Sana08}, and \citealt{debeck06}). However, the number of
orbital solutions for massive binaries is still relatively small. In
our previous works documenting the number and statistics of massive
binaries in Cygnus OB2 \citep{Kiminki07, Kiminki08, Kiminki09}, we
discovered nine systems (MT059, MT145, MT252, MT258, MT372, MT720,
MT771, Schulte~3, and Schulte~73; notation from \citealt{MT91} and
\citealt{Schulte56}). We provided spectroscopic solutions to these
nine and two additional systems (A36 and A45; notation from
\citealt{Comeron02}) as part of our ongoing radial velocity survey in
one of the Galaxy's super OB clusters. The primary objective of the
Cyg~OB2 radial velocity survey is to aid our understanding of numerous
open questions involving massive stars, including their formation, the
predicted rates of supernovae type Ib/c arising through binary
channels, and whether there is a correlation between cluster density
and the binary fraction therein.

In this work, we report discovery of two new double-lined
spectroscopic binaries (SB2s), and we provide full orbital solutions
(periods, mass ratios, eccentricities and probable inclinations) for
these and four known binaries previously lacking full solutions.  
This brings the number of massive binaries having full or partial 
solutions to 20, the most of any single cluster or association.  
In a companion paper \citep{Kiminki11b}, we summarize the current state of
massive binaries in the literature and use the combined binary orbital
parameter information and ephemerides for 113 massive stars in Cyg~OB2
to infer the intrinsic orbital parameter distributions and the binary
fraction of the Association.  We present these findings as a
complement to analyses of massive stars in other clusters (such as the
work with Sco~OB2 by \citealt{Kouwenhoven07}, NGC~6231 by
\citealt{Sana08}, NGC~2244 by \citealt{Mahy09}, and NGC~6611 by
\citealt{Sana09}). For a more comprehensive discussion of the scope of
this project see \citet{Kiminki11b}.

Section~2 reviews the observations of this ongoing 12-year radial
velocity survey and discusses the method we used to determine radial
velocities. Section~3 discusses the method for computing orbital
solutions and provides new spectroscopic orbital solutions for MT311,
MT429, MT605, MT696, MT720, and MT771. Finally, Section~4 summarizes
the survey findings to date.

\section{Observations and Radial Velocity Measurements}
Papers I--III (\citealt{Kiminki07}, \citealt{Kiminki08}, and
\citealt{Kiminki09} respectively) detail the observations of this
survey through 2008 October (see Table~1 of Paper III).  Since then,
we have obtained eleven additional datasets with the WIRO-Longslit
spectrograph on the Wyoming Infrared Observatory (WIRO) 2.3~m
telescope. The observation dates, including Heliocentric Julian
Dates, for these datasets are listed in Table~\ref{obs4.tab}. We used
the 1800 l~mm$^{-1}$ grating in first order and obtained a typical
spectral resolution of $\sim$1.5~\AA\ FWHM across the chip. These
observations took place over 26 nights between 2008 December 17 and
2010 October 27 to primarily examine the H$\alpha$, \ion{He}{1}, and
\ion{He}{2} absorption lines in suspected SB2s. Exposure times for all
observations varied from 300~s to 4500~s (in multiples of 300--900~s)
depending on weather conditions and yielded a maximum signal-to-noise
ratio (S/N) of 200:1 for the brightest stars. The spectral coverage
was 5250--6750~\AA. Copper-Argon lamp exposures were taken after each
star exposure to wavelength calibrate the spectra to an rms of
0.03~\AA\ (1.4~\kms\ at 6400~\AA). Spectra were Doppler corrected to
the heliocentric frame. All new datasets were reduced using standard
IRAF\footnote{IRAF is distributed by the National Optical Astronomy
Observatories, which are operated by the Association of Universities
for Research in Astronomy, Inc., under cooperative agreement with the
National Science Foundation.} reduction routines as outlined in
Paper~I.

We obtained radial velocities for the single-lined spectroscopic
binary (SB1), MT429 via the IRAF cross-correlation task XCSAO in the
RVSAO package \citep{xcsao}, using a model stellar atmosphere
\citep[TLUSTY]{LHub2003} of the appropriate effective temperature and
gravity (Paper I). Wavelength regions used include
$\lambda\lambda$4020--4500~\AA\ (excluding the DIB at
$\lambda$4428~\AA) for the WIYN spectra,
$\lambda\lambda$5250--6750~\AA\ (excluding all regions except
\ion{He}{2}~$\lambda$5411~\AA, \ion{He}{1}~$\lambda\lambda$5876,
6678~\AA, and H$\alpha$) for the 2008 and 2009 WIRO data, and
$\lambda\lambda$4200--6000~\AA\ (excluding the DIB at
$\lambda$4428~\AA) for the 2010 WIRO data. \citet{xcsao} calculate
the uncertainties within the XCSAO task as

\begin{equation}
\sigma_v = \frac{3w}{8(1+r)},
\end{equation}

\noindent a function of the $r$ statistic (the ratio of the
correlation peak height to the amplitude of antisymmetric noise)
derived in \citet{tonry79}. $w$ in this equation is the FWHM of the
correlation peak.

In order to measure radial velocities for the SB2s, we fit multiple
Gaussians to the \ion{He}{1} $\lambda\lambda$5876, 6678~\AA\ and
H$\alpha$ lines. We obtain rough fits to the most deblended (and
highest S/N) spectra using the SPLOT routine in IRAF. These values are
then used to fit two four-parameter Gaussians simultaneously to those
same spectra using the IDL multiple Gaussian fitting routine
ARM\_MULTGAUSS (written by Andrew R. Marble at the University of
Arizona), which makes use of the non-linear, least-squares, curve
fitting IDL package MPFIT \citep{Markwardt08}. We then adopt the
average of the Gaussian widths and depths as fixed parameters to fit
line centers for all deblended spectra of the system. The radial
velocities are then calculated from the best-fit Gaussian
centers. Gaussian parameter uncertainties are obtained from the
square-root of the diagonal elements of the covariance matrix
multiplied by the measured $\chi^2$ of the fit.  The overall radial
velocity uncertainties are obtained by summing the uncertainties on
the line centers in quadrature with the typical uncertainty in the
wavelength calibration (1.4~\kms\ for WIRO and 2~\kms\ for WIYN).  The
wavelength solutions of the WIRO-Longslit spectra have been adjusted
to a common zero point using the mean of the measured radial
velocities of the interstellar \ion{Na}{1}~$\lambda\lambda$5889.95,
5895.92~\AA\ lines, which we assume are invariant for a given star.
WIYN and Keck spectra have been calibrated using radial velocity
standard stars \citep{Kiminki07,Kiminki08}.

\section{Spectroscopic Solutions} 
To estimate the periods from the radial velocity data for both the
SB1s and SB2s, we analyze the power spectrum as filtered through the
the CLEAN deconvolution algorithm of \citet{Roberts87}.  For
computation of the orbital solutions we used the Binary Star Combined
Solution Package (referred to as BSCSP in this work) of
\citet{Gudehus01} which makes use of a nonlinear, least-squares
subroutine (NONLINFIT) to solve for orbital parameters of binary (SB1
and SB2) and multiple systems using spectroscopic and/or astrometric
data. As an added check, we also computed single-component
spectroscopic solutions with the least-squares, curve fitting program
SBCM \citep{Morbey74} and combined component solutions with SBTWOS,
our own MATLAB-based orbital solution program utilizing Markov-Chain
Monte Carlo techniques. Solutions from all programs were in good
agreement, but BSCSP and SBTWOS yielded orbital parameter
uncertainties up to three times larger than those from SBCM. In this
work, we present the results with the more conservative uncertainties
from BSCSP. For the eclipsing, spectroscopic, triple system MT429, we
also used PHOEBE \citep{PRSA05}, an eclipsing binary modeling program
based on the code of \citet{Wilson71}. PHOEBE provides orbital
solutions for eclipsing binary systems with photometric and/or
spectroscopic data.

\subsection{MT311}
\begin{figure}
\centering
\epsscale{0.85}
\plotone{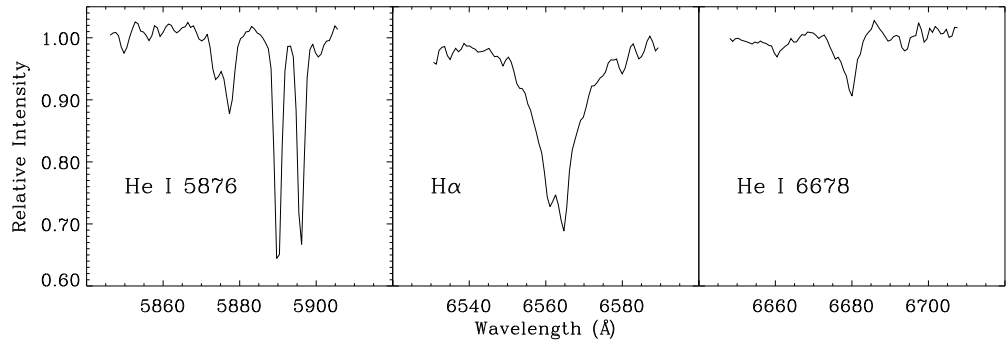}
\caption{A composite of four spectra between a phase of $\phi=0.457$
and $\phi=0.655$ for MT311. The three panels show \ion{He}{1}~$\lambda$5876~\AA,
H$\alpha$, and \ion{He}{1}~$\lambda$6678~\AA\ at quadrature respectively.
\label{311quad}}
\end{figure}

\begin{figure}
\centering
\epsscale{0.85}
\plotone{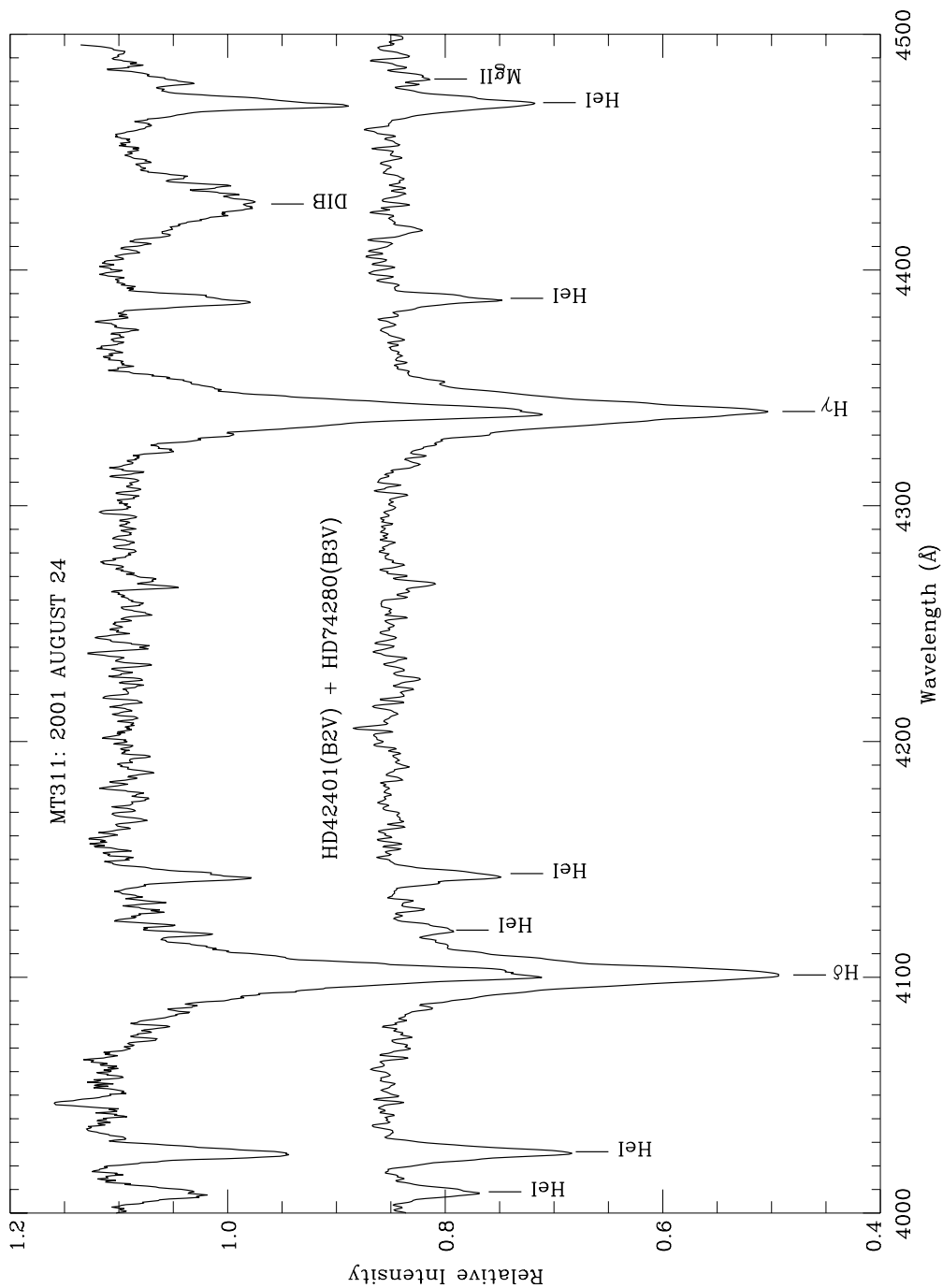}
\caption{Spectrum of MT311 (top) obtained on 2001 August 24 and a composite
of two spectra that best match the spectral types of the primary and 
secondary from the \citet{WF90} digital atlas shifted to the appropriate 
radial velocities.
\label{twoplot}}
\end{figure}

At V$\sim$14, the SB2 MT311 is among the faintest in our sample, and
the resulting spectra have an average signal-to-noise ratio (S/N) of
$\sim$60. The primary is classified a B2V by \citet{MT91}. Given the
lower S/N and velocity separation of $\sim$210~\kms, we can only
determine approximate spectral types for the components.  An absence
of \ion{He}{2} and relatively strong \ion{He}{1} places both
components in the early B range. The 2001 August 24 spectrum, obtained
at WIYN, shows weak \ion{Mg}{2}~$\lambda$4481 \AA\ and absent or weak
silicon, nitrogen, and carbon (i.e., \ion{C}{3}~$\lambda$4070 \AA,
\ion{C}{2}~$\lambda$4267 \AA, \ion{Si}{4}~$\lambda\lambda$4089,4116
\AA, \ion{Si}{2}~$\lambda\lambda$4128,4130 \AA,
\ion{O}{2}~$\lambda\lambda$4415,4417 \AA, \ion{N}{2}~$\lambda$3995
\AA). This suggests that both components are likely still on the main
sequence. The temperature-sensitive ratio
\ion{He}{1}~$\lambda$4471~\AA:\ion{Mg}{2}~$\lambda$4481~\AA\ indicates
an approximate temperature class of B2--B3 for both
components. Figure~\ref{311quad} shows the similarity between the line
depths for \ion{He}{1}~$\lambda$5876~\AA, H$\alpha$, and
\ion{He}{1}~$\lambda$6678~\AA, suggesting similar luminosities and
providing additional support for comparable temperature classes. We,
therefore, tentatively classify the components a B2V and a B3V. In
Figure~\ref{twoplot}, we show the 2001 August 24 spectrum of MT311
obtained at WIYN (top).  We provide a composite spectrum below MT311,
composed of HD42401 (B2V) and HD74280 (B3V) from \citet{WF90}. HD42401
and HD74280 have been shifted to the appropriate component radial
velocities from Table~\ref{OC24} and scaled to the appropriate
luminosity ratio for a B2V and B3V. The MT311 spectrum is smoothed to
approximately the resolution of the \citet{WF90} spectra.  The
stronger absorption present in the \ion{He}{1}~$\lambda$4471~\AA\ and
\ion{Mg}{2}~$\lambda$4481~\AA\ lines in the MT311 spectrum may be a
normalization effect given that the lines fall near the end of the
chip. The remaining lines are consistent with the composite spectrum.

\begin{figure}
\epsscale{0.9}
\centering
\plotone{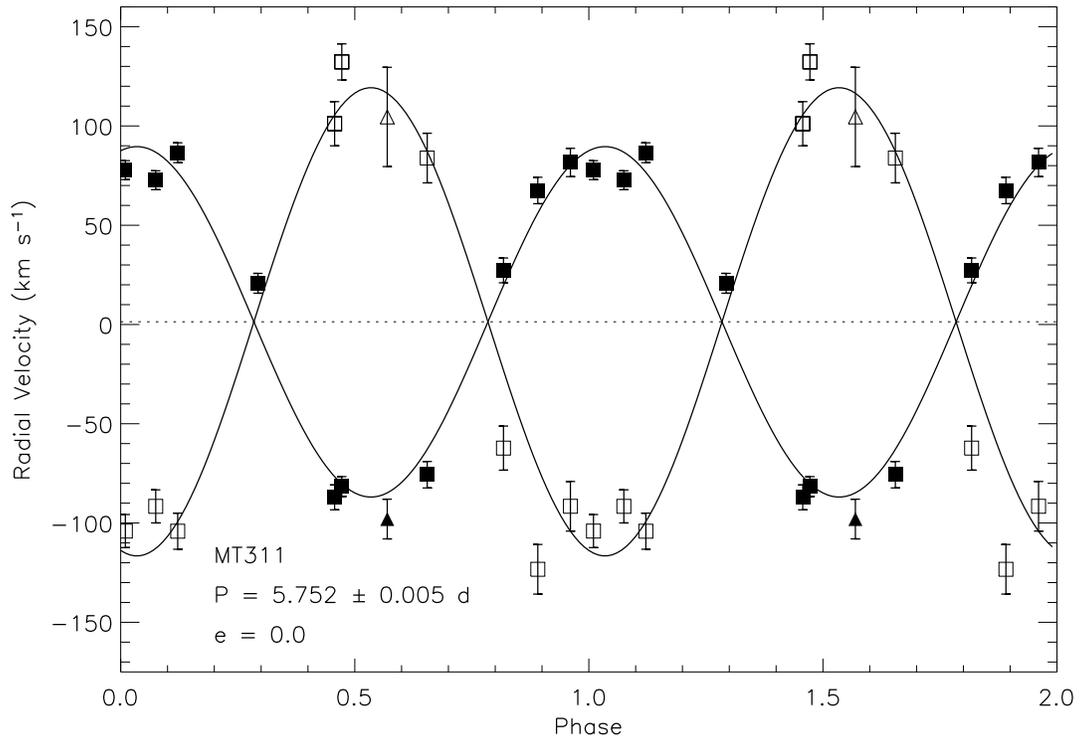}
\caption{Heliocentric radial velocity curve and orbital solution for
MT311 using 11 of the highest S/N spectra. The filled points
correspond to the primary (B2V) and the open points correspond to the
secondary (B3V). The triangle represents an observation taken with
Hydra at WIYN and the squares represent observations taken with
WIRO-Longslit at WIRO.
\label{311curve}}
\end{figure}

Table~\ref{OC24} contains the ephemeris for this system, including
radial velocities for the 11 highest S/N spectra. The majority of the
spectra were obtained from 2009 June 27 through 2009 September 4 with
the Wyoming Infrared Observatory (WIRO) Longslit spectrograph. One
observation is from 2001 with the WIYN\footnote{The WIYN Observatory
is a joint facility of the University of Wisconsin, Indiana
University, Yale University, and the National Optical Astronomy
Observatory.} 3.5~m telescope and Hydra spectrograph. With these 11
spectra, covering a span of eight years, we calculate radial
velocities from the \ion{He}{1}~$\lambda$5876~\AA\ line
(\ion{He}{1}~$\lambda$4471~\AA\ line for the WIYN spectrum) and obtain
a period of 5.75~days from the power spectrum for both
components with no significant aliases. The best orbital solution,
displayed in Figure~\ref{311curve}, yields a period of
$P=5.752\pm0.005$~days with a fixed eccentricity of zero. Allowing the
eccentricity to vary results in an insignificant decrease in the rms
of the fit and yields $e=0.22\pm0.05$. 
Given that the 11 data points do not provide a strong 
constraint on the eccentricity, we adopt the solution
with the fixed zero eccentricity. In the figure, solid symbols
correspond to the radial velocities of the primary, open symbols
correspond to the radial velocities of the secondary, and the
different shapes represent the different instruments/observatories
(i.e, squares represent observations taken with WIRO-Longslit and
triangles represent observations taken with Hydra at WIYN). We found
it more difficult to fit the secondary's spectral features owing to
the low S/N and heavy blending of profiles and thus only utilized 10
measurements (the measurement at $\phi=0.808$ was excluded). With a
period of $P=5.752\pm0.005$~days, an eccentricity of $e=0.0$, and
semi-amplitudes of $K_{1}=88\pm5$~\kms\ and $K_{2}=118\pm9$~\kms, we
calculate masses of $M_{1}\mathrm{sin}^{3}i=3.0\pm0.7$~\msun\ and
$M_{2}\mathrm{sin}^{3}i=2.2\pm0.4$~\msun, and semimajor axes of
$a_{1}\mathrm{sin}i=10.0\pm0.6$~\rsun\ and
$a_{2}\mathrm{sin}i=13.40\pm0.01$~\rsun. This yields a mass ratio of
$q=0.75\pm0.07$ and is consistent with the theoretical mass ratio of
$q=0.76$ for a B2V$+$B3V system \citep{drilling}. The calculated lower
limit masses indicate that the system needs to be inclined at
$i\sim$42\deg\ to be consistent with the theoretical masses of
10~\msun\ and 7.6~\msun\ for a B2V star and B3V star \citep{drilling}.

The computed heliocentric systemic velocity for MT311 is
$\gamma=1\pm4$~\kms. This differs by nearly 11~\kms from the average
systemic velocity of $-10.3$~\kms\ for other members of the
Association (Paper~I), calling into question its membership. 
If we adopt an apparent $V$ magnitude of 13.87
\citep{MT91} for the system, the B-V color of 1.39 for the system
\citep{MT91}, an absolute $V$ magnitudes of -2.45 and -1.94 for a B2V
and B3V respectively \citep{HM84}, an intrinsic color $(B-V)_0=-0.21$
and a reddening law with $R_V$=3.0, the resulting spectrophotometric
distance to MT311 is $\sim$2.6~kpc (distance modulus of 12.07). The
accepted distance to Cyg~OB2 is near 1.7~kpc \citep{Hanson03} but with
a relatively large uncertainty owing to the uncertainty on absolute
magnitudes for massive stars and the observed variability in the
total-to-selective extinction, $R_V$ \citep{Patriarchi2003}.  Although
our computed spectrophotometric distance is 50\% larger than the
accepted distance to Cyg~OB2, the uncertainties are sufficiently large
that we cannot draw a definitive conclusion regarding
the membership of MT311.

Table~\ref{OC24} provides the ephemerides for MT311, MT605, MT696,
MT720, and MT771, listing the date (HJD$-$2,400,000), phase ($\phi$),
measured radial velocities ($V_{r1}$ and $V_{r2}$), and observed minus
calculated velocities ($O_{1}-C_{1}$ and $O_{2}-C_{2}$). Error
estimates are shown in parentheses. The complete list of orbital
elements and physical parameters for MT311, MT429, MT605, MT696,
MT720, and MT771 appears in Table~\ref{orbparms4.tab}. Listed within
the table are the period in days ($P$), eccentricity of the orbit
($e$), longitude of periastron in degrees ($\omega$), systemic radial
velocity ($\gamma$), epoch of periastron ($T_0$), primary and
secondary semi-amplitudes ($K_1$ \&\ $K_2$), adopted or calculated
minimum primary and secondary masses in solar masses ($M_1$ \&\
$M_2$), primary and secondary mass functions in solar masses ($f(m)_1$
\&\ $f(m)_2$), spectral classifications from this survey (S.C.$_1$ \&\
S.C.$_2$), the minimum primary and secondary semi-major axes in solar
radii ($a_1$sin~$i$ \&\ $a_2$sin~$i$), and finally, the rms of the
best fitting orbital solution 
($rms_1$ \&\ $rms_2$). Additionally, the inclination, primary
and secondary effective temperatures, and component stellar radii are
included for MT429. 

\subsection{MT429}
\citet{MT91} classify MT429 ($\equiv$V2186~Cyg) 
as a B0V. \citet{PK98} showed that this is
an eclipsing system of the Algol type, estimating a period of 2.9788
days based on an incomplete light curve.  Our spectra
of MT429 initially suggested it to be an SB1 with low-amplitude 
radial velocity variations at a very similar period.  

\begin{figure}
\centering
\epsscale{0.9}
\plotone{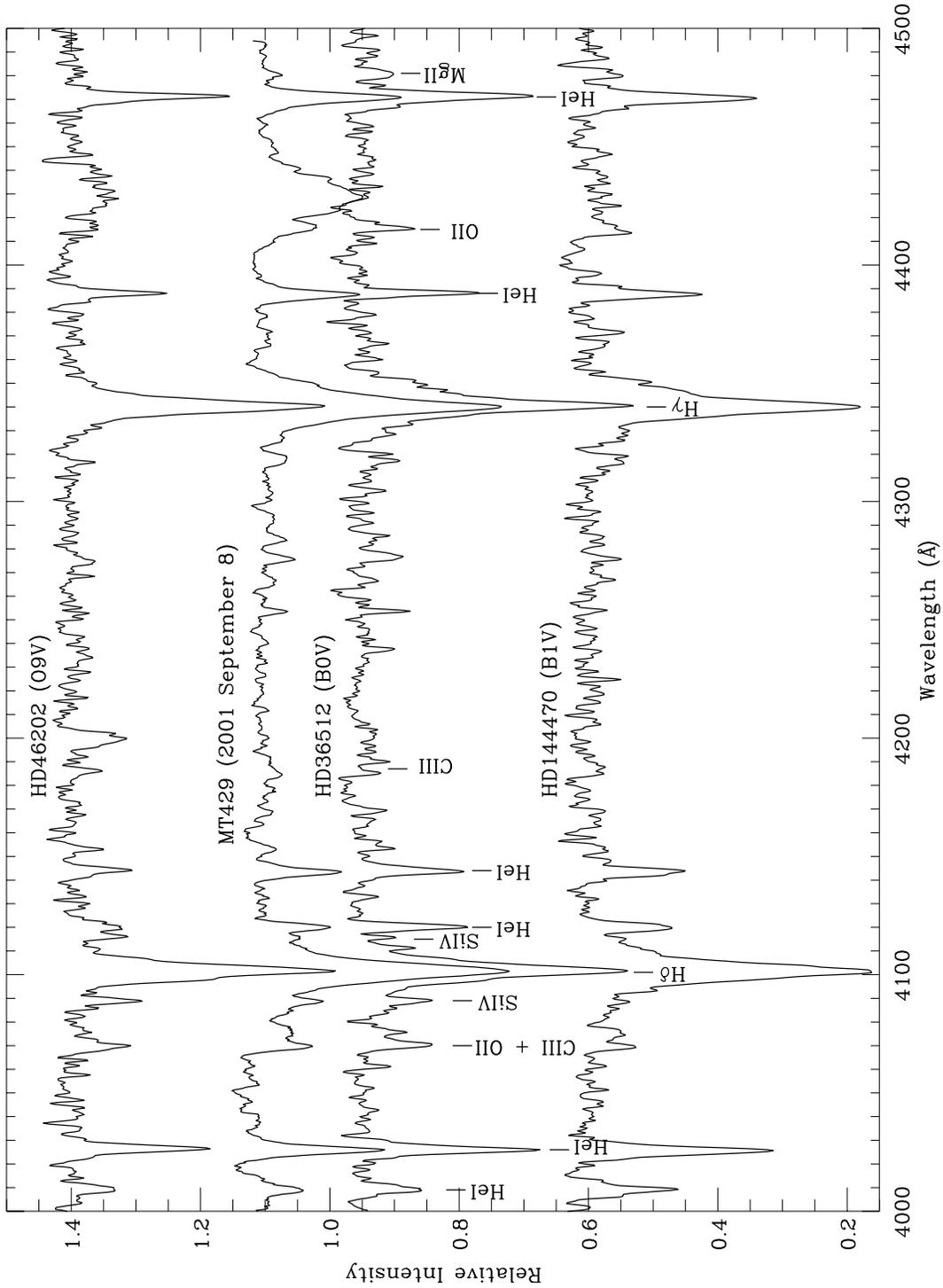}
\caption{Spectrum of MT429a (second from top) obtained on 2001
September 8 and an O9V (HD46202), a B0V (HD36512), and a B1V
(HD144470) from the \citet{WF90} digital atlas for comparison.
\label{429comp}}
\end{figure}

Figure~\ref{429comp} shows a slightly smoothed, high
quality spectrum of MT429, obtained 2001 September 8, with O9V
(HD46202), B0V (HD36512), and B1V (HD144470) spectra from the digital
atlas of \citet{WF90} for comparison. With the exception of the lower
S/N 2000 September 18 Keck spectrum, our spectra of MT429 do not
cover the important \ion{Si}{3}~$\lambda$4552 \AA\ line, used with the
\ion{Si}{4}~$\lambda$4089 \AA\ line and the
\ion{Si}{2}~$\lambda\lambda$4128--4130~\AA\ lines as the primary
temperature class indicator in early B stars. However, the WIYN 
spectrum in the figure shows a moderately weak
\ion{Si}{4}~$\lambda$4089 \AA\ line and a moderately strong
\ion{C}{3}~$+$~\ion{O}{2} blend at $\lambda\lambda$4070 \AA. These lines
quickly diminish with decreasing temperature in early B stars
\citep{WF90}. There is also no evidence for \ion{He}{2}~$\lambda$4200.
This agrees with a temperature class of B0--B1. The primary luminosity
indicator (\ion{Si}{4}~$\lambda\lambda$4089,4116~\AA\ to the nearby
\ion{He}{1} lines) also agrees with a V luminosity class. All of our
spectra support the classification of MT429 as B0V, within a fraction
of a temperature class, consistent with the \citet{MT91}
classification.

We obtained 16 spectra on MT429 from 2000 September 18 through 2010
October 27, including one at Keck, five at WIRO, and 10 at WIYN. The
WIYN spectra were cross-correlated using a wavelength range of
$\lambda\lambda$4020--4500~\AA\ (excluding the DIB at
$\lambda$4428~\AA). The WIRO 2010 spectra were cross-correlated using
a wavelength range of $\lambda\lambda$4200--6000~\AA\ (excluding the
DIB at $\lambda$4428~\AA\ and the interstellar (IS) \ion{Na}{1} lines
at $\lambda\lambda$5890,5895~\AA). The 2008--2009 WIRO spectra were
cross-correlated using a wavelength range of
$\lambda\lambda$5400--6700~\AA\ (excluding all but the \ion{He}{1},
\ion{He}{2}, and hydrogen lines owing to the large number of IS lines
and DIBs in this spectral range). The Keck spectrum was
cross-correlated using a wavelength range of
$\lambda\lambda$4050--5200~\AA\ (excluding the DIB at
$\lambda$4428~\AA)

The strongest peak in the power spectrum corresponds to a period of
$P=2.9786$~days with no significant aliases, consistent with the
photometric periods of \citet{PK98} and
\citet{Henderson2011}. Allowing all orbital parameters to vary, BSCSP
provides an orbital fit with $P=2.9785$, $e=0.24$, $K=17.7$~\kms, and
rms of 3.0~\kms. We excluded one outlier from the fit, the 2008 June
15 WIYN measurement ($V_r=-48$~\kms). Fixing the eccentricity at zero
provides a comparable fit to the data with a similar rms of 
(3.4~\kms), period of $P=2.9786\pm0.0001$~days and systemic velocity
of $\gamma=-14\pm1$~\kms.  The resulting mass function of
$f(m)=0.0012\pm0.0003$ would suggest that either the secondary is a
low-mass star ($M_{1}\sim0.8$~\msun) or that the inclination is very
small.  A small inclination is inconsistent with the observed
eclipses.  A large inclination coupled with a low-mass companion could
conceivably explain the observed radial velocity curve, but such a
scenario entails that only primary eclipses would be observed given
the vast difference in temperature between a main sequence or evolved
$\sim$1~\msun secondary and a B0V primary.  However, existing
photometric data clearly show both a primary and a secondary eclipse.

\begin{figure}
\centering
\epsscale{0.9}
\plotone{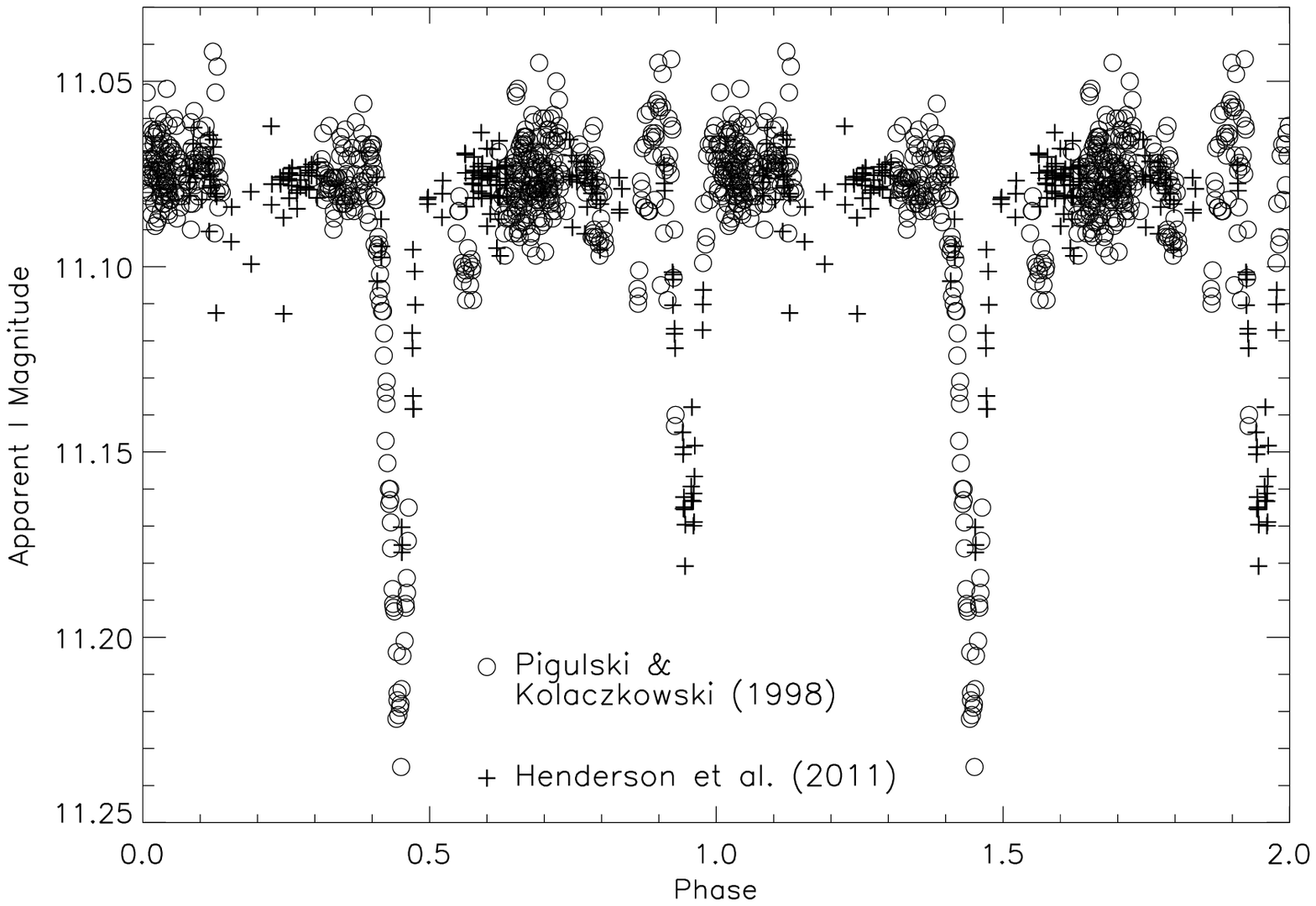}
\caption{Combined I-band photometric data from \citet{PK98} and
\citet{Henderson2011} folded at a period of 2.9786~days.
\label{429only}}
\end{figure}

We combined the photometric data of \citet{PK98} with that of
\citet{Henderson2011} who also obtained I-band photometry, finding a
nearly identical period of 2.9788~days. We adopted an apparent I
magnitude of 11.08 based on the mean maximum magnitude from
\citet{Henderson2011}. There are 3,710 days between the end of the
\citet{PK98} data and the start of the \citet{Henderson2011}
data. This necessitates shifting the \citet{Henderson2011} data by
phase $\phi=0.225$. This is approximately equal to a 0.0002~day
difference in the calculated period but is consistent within the
calculated period uncertainty.\footnote{The radial velocity data also
required a phase shift amounting to 0.09 after 1,137~days, which is
equivalent to a period difference of $\sim$0.0002~days and consistent
within our calculated orbital period uncertainty.}
Figure~\ref{429only} shows the combined, adjusted photometric data
folded at the 2.9786~day spectroscopic period.  Open circles denote
the \citet{PK98} data, and the crosses denote the
\citet{Henderson2011} data.  The data show a primary to secondary
eclipse depth of slightly less than 2:1, indicating components with
similar temperatures.  Adopting the period, eccentricity, epoch of
periastron, and longitude of periastron from the spectroscopic
solution, the data were best fit with an inclination of
$i=74.3^\circ$, mass ratio of $q=0.59$, separation of
$a\mathrm{sin}i=22.5$~\rsun, primary radius of $R_{1}=5.5$~\rsun,
secondary radius of $R_{2}=3.9$~\rsun, primary temperature of
$T_{1}=20,900$~K, and secondary temperature of $T_{2}=15,200$~K. These
results are consistent with a B2V primary and B5V secondary
\citet{drilling}. However, the semiamplitude of 16~\kms\ measured from the
spectroscopic solution does not agree with the predicted 137~\kms\
primary radial velocity semiamplitude from PHOEBE.

\begin{figure}
\centering
\epsscale{0.9}
\plotone{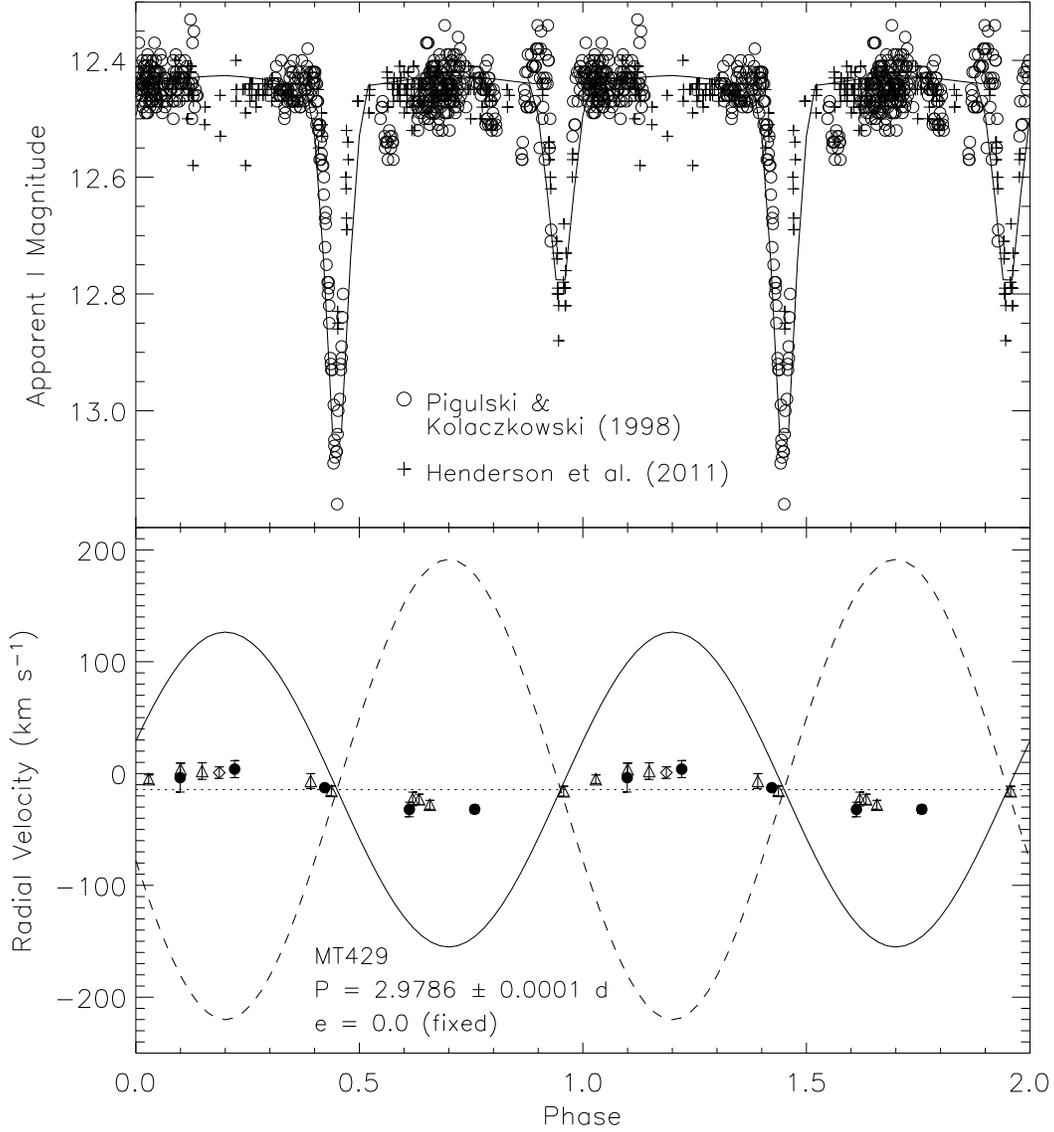}
\caption{Upper panel: the I-band photometric data and best-fit PHOEBE
light curve solution for MT429.  Open circles are the I magnitudes
obtained from \citet{PK98}, and the crosses are the I magnitudes
obtained from \citet{Henderson2011} Lower panel: theoretical velocity
curve for the putative B3V (solid curve) and B6V (dashed curve)
components under the triple system (B0V$+$B3V$+$B6V) scenario
described in the text for MT429. Data points (filled circles for WIRO,
triangles for WIYN, a diamond for Keck) show the apparent observed
heliocentric radial velocity in the composite spectrum which is
dominated by the (presumably constant velocity) B0V modulated by the
$\sim$140 \kms\ amplitude signal of the fainter B2V.
\label{429fit2}}
\end{figure}

The seemingly inconsistent conclusions from the light curve analysis
(a B2V + B5V eclipsing binary) and the spectroscopic velocity curve
analysis (a B0V with a velocity amplitude of 16 \kms\ implying a very
low-mass companion) posed a conundrum until the the discovery that
MT429 is a close visual double. \citet{saida11} present
adaptive-optics K-band images and $HST$ fine guidance sensor
measurements showing that the MT429 system is a visual double with a
separation of 0$\farcs$08 (138 AU projected separation at a distance
of 1.7 kpc). The fainter component, MT429b, is 1.05 mag fainter at V
band than MT429a, consistent with an early-to-mid B star and a B0V at
a common distance and reddening.  {\it A consistent picture of MT429
emerges if it is actually a triple system consisting of a B0V star
(MT429a) separated by at least 138 AU from a tight B2V+B5V binary
(MT429b+MT429c) in a 2.9786 day orbit.}  Statistically, the small
angular separation of 0$\farcs$08 between the two dominant components
in MT429 is improbable unless there is a physical association
\citep{saida11}. With this new information, we adjusted the
photometric data to remove the additional flux contributed by the B0V,
estimated to be $\sim$72\%\ of the total flux by
\citet{saida11}. Because a larger percentage of the total light is
being eclipsed with the B0V flux subtracted, the eclipse depths appear
deeper compared to the light curve in Figure~\ref{429only}. The upper
panel in Figure~\ref{429fit2} shows the adjusted, combined photometric
data with the new PHOEBE fit. Again, adopting the period,
eccentricity, epoch of periastron, and longitude of periastron from
the spectroscopic solution, the best fit to the data results in an
inclination of $i=89^\circ$, mass ratio of $q=0.68$, separation of
$a\mathrm{sin}i=20.4$~\rsun, primary radius of $R_{1}=4.8$~\rsun,
secondary radius of $R_{2}=3.6$~\rsun, primary temperature of
$T_{1}=19,100$~K, and secondary temperature of $T_{2}=13,700$~K. This
is most consistent with a B3V primary and B6V secondary
\citep{drilling}. The new result predicts a radial velocity primary
semiamplitude of $\sim$140~\kms\ (205~\kms\ for the secondary), still
in sharp disagreement with the measured semiamplitude of 16~\kms.  The
lower panel of Figure~\ref{429fit2} shows the theoretical velocity
curves of the B3V (solid curve) and B6V (dashed curve) stars
prescribed by the above scenario. We believe the answer to this
discrepancy lies in the fact that the B3V is roughly 5.5$\times$
fainter than the B0V and the B6V is significantly fainter than the B3V
in the visual band, and the large velocity amplitude of the tight
binary is masked (spectral dilution) by the much brighter (apparently)
constant-velocity B0V component MT429a. Simulations wherein we combine
spectra with appropriate flux and signal-to-noise ratios support the
conclusion that a 140 \kms\ velocity amplitude signal from MT429b is
diluted to the level of $\sim$16 \kms\ in a composite B0V+B3V
spectrum. The symbols in the lower panel (filled circles for WIRO,
triangles for WIYN, a diamond for Keck) show the apparent velocity
curve measured in the composite spectrum of the system. The dotted
line shows the systemic velocity of the binary
($\gamma=-14\pm1$). While the period of the folded velocity data
agrees well with the photometric data, the small implied velocity
amplitude of 16 \kms\ is not representative of the velocity of any
component within the system. MT429, then, is the first massive triple
system discovered in the CygOB2 radial velocity survey. It joins
Schulte~5 and Schulte~8 as a growing group of N$>$2 multiple-star
systems in the Cyg OB2 Association.

We adopt from BSCSP the values and uncertainties pertaining to the
orbital period, eccentricity, epoch of periastron, and angle of
periastron for the B3V+B6V pair and list these in
Table~\ref{orbparms4.tab}.  We adopt the values for component radii,
masses, effective temperatures, theoretical semiamplitude semimajor
axes, and mass ratio from PHOEBE. These pertain to the tight B3V+B6V
system.  The listed systemic velocity of the system pertains to the
composite system.  The RV ephemeris for MT429 is listed in
Table~\ref{OC34} and includes the date (\textit{$HJD-2,400,000$}),
phase (\textit{$\phi$}), measured radial velocity (\textit{$V_r$ }),
cross-correlation $1\sigma$ error (\textit{$1\sigma$~err}).  We
reiterate here that, while these velocities show excellent agreement
with the period and phase angle derived from the light curve, the {\it
amplitudes} are not physically meaningful, as they reflect the radial
motion of the B3V component as diluted by the (apparently constant
velocity) spectrum of the brighter B0V component.

\subsection{MT605}
\begin{figure}
\centering
\epsscale{0.85}
\plotone{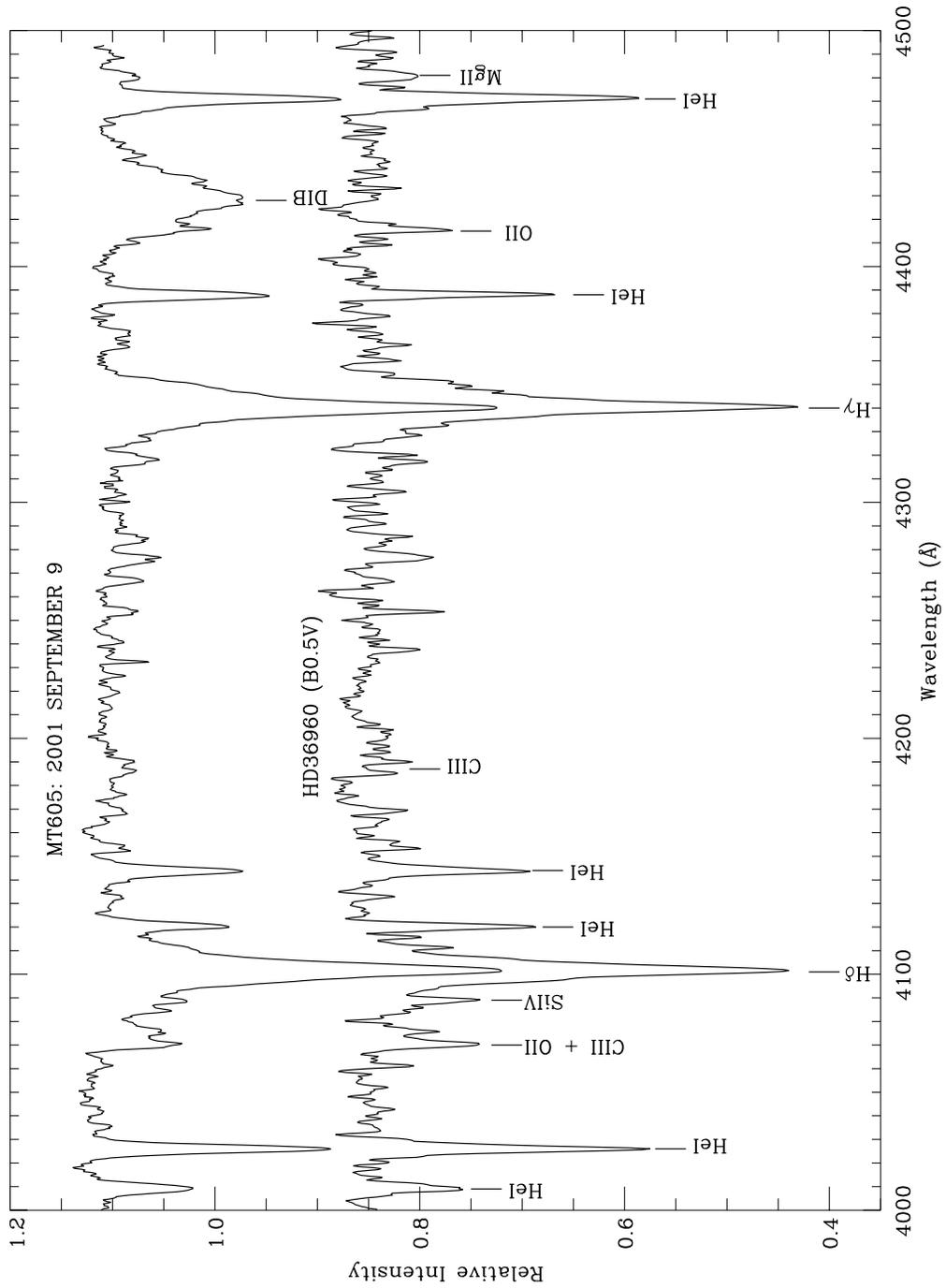}
\caption{Spectrum of MT605 near quadrature (top) obtained on 2001
September 9 and HD36960 (B0.5V) from the \citet{WF90} digital atlas
for comparison.
\label{mt605comp}}
\end{figure}

\citet{MT91} classify MT605 a B0.5V. In Paper~I, we typed the presumed
single system a B1V based on visual comparisons with spectra from the
\citet{WF90} digital atlas. We now know that this is an
SB2. Examination of the highest quality spectra for this system shows
that neither component's spectrum displays any \ion{He}{2}, indicating
that the components are likely later than B0. However, very weak
\ion{Mg}{2}~$\lambda$4481~\AA\ absorption relative to
\ion{He}{1}~$\lambda$4471~\AA\ indicates the components are earlier
than B2. Furthermore, because \ion{Si}{3}~$\lambda$4552~\AA\ and
\ion{Si}{4}~$\lambda$4089~\AA\ are also weak and
\ion{C}{3}$+$\ion{O}{2}~$\lambda\lambda$4070, 4650~\AA\ absorption is
strong, we reclassify the primary a B0.5V \citep[in agreement with the
original assessment of ][]{MT91} with acknowledgment that the small
velocity separation ($\sim$160~\kms\ for \ion{He}{1}) at a resolution
of R$\sim$4500 makes the luminosity classification difficult and still
somewhat uncertain. With a mass ratio near unity and a luminosity
ratio also near unity (determined from examination of all lines while
in or near quadrature), we also tentatively classify the secondary a
B0.5--B1 with an indeterminate luminosity
class. Figure~\ref{mt605comp} displays the 2001 September 9 spectrum
of MT605 taken near quadrature (top) and obtained at WIYN and HD36960
(B0.5V) from the \citet{WF90} digital atlas for comparison. The WIYN
spectrum has been boxcar smoothed to approximately the resolution of
the \citet{WF90} spectra. The comparison spectrum is in good agreement
with the WIYN spectrum of MT605, with the exception of the
\ion{C}{3}$+$\ion{O}{2}~$\lambda\lambda$4070,4650~\AA\ blend and
\ion{Si}{4}~$\lambda$4089~\AA\ absorptions, which appear slightly
stronger in HD36960.

\begin{figure}
\centering
\epsscale{0.9}
\plotone{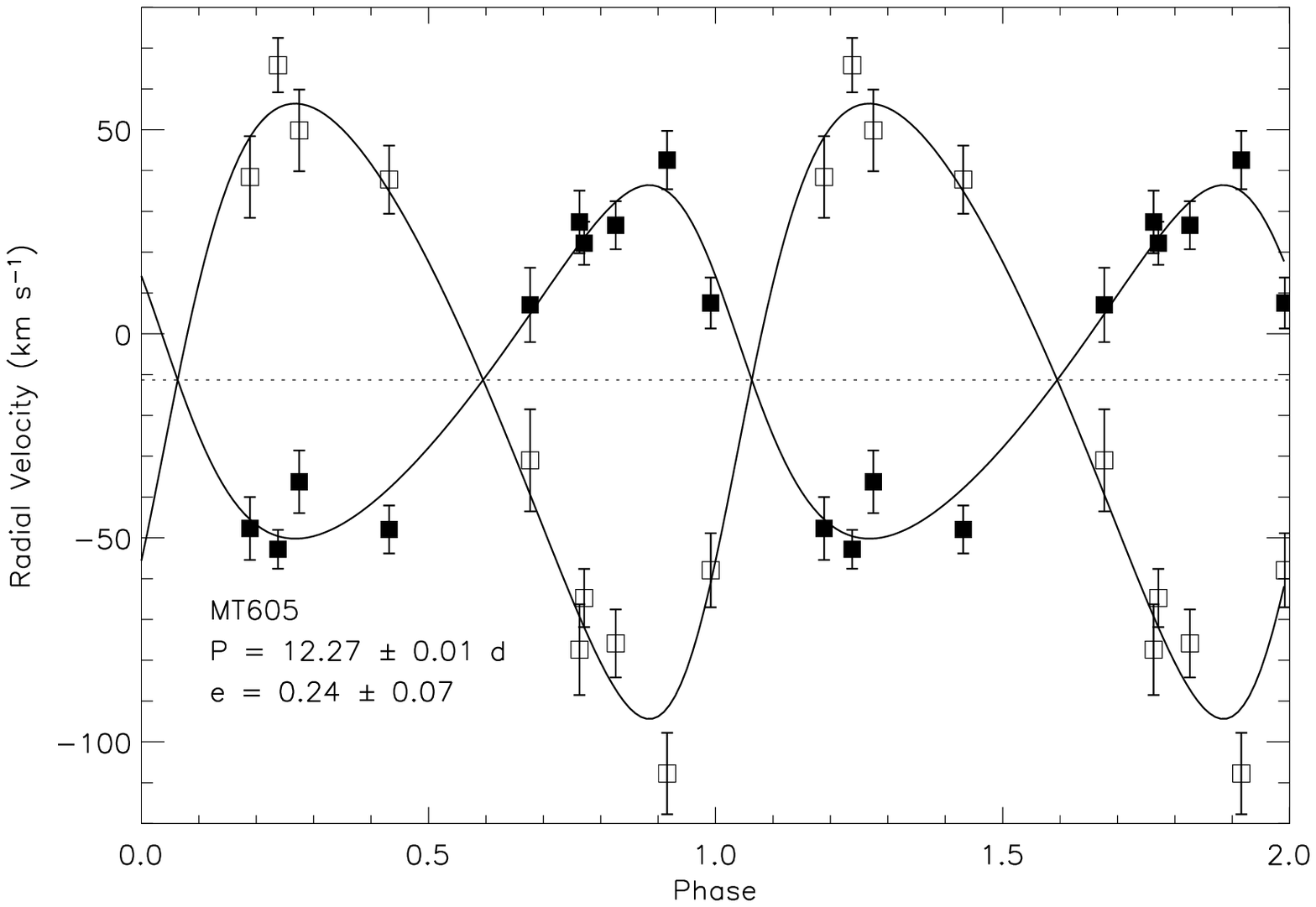}
\caption{Heliocentric radial velocity curve and orbital solution for MT605 using 10
of the highest S/N spectra obtained at WIRO with WIRO-Longslit.
\label{mt605curve}}
\end{figure}

We observed this system 37 times between 1999 and 2009, obtaining 10
spectra (all with WIRO-Longslit) with sufficiently high enough S/N and
large enough component profile velocity separations ($\sim$150~\kms\
at max) to measure radial velocities. Because of the small number of
suitable spectra available for measuring radial velocities,
determining an estimate of the period from the power spectrum is
difficult. We, therefore, fit single Gaussians to the
\ion{He}{1}~$\lambda$5876~\AA\ and H$\alpha$ lines and measured the
profiles' Gaussian FWHM in 18 of the WIRO spectra (17 for
\ion{He}{1}~$\lambda$5876~\AA). From these measurements, we observe a
strong peak in the power spectrum at $\sim$6~days for both \ion{He}{1}
and H$\alpha$. This represents half the period of the binary and tells
us the location of the primary and secondary spectral features when
the profiles are blended. Using this information, we performed the
same technique as with the other SB2s to calculate radial velocities
from the \ion{He}{1}~$\lambda$5876~\AA\ line for the 10 spectra. The
power spectrum revealed that the likely period from the 10 spectra is
12.26 days for both the primary and secondary star spectral features.
The best orbital solution provided by BSCSP and SBTWOS yields a period
of $P=12.27\pm0.02$~days, an eccentricity of $e=0.24\pm0.07$, and
semi-amplitudes of $K_{1}=44\pm3$~\kms\ and $K_{2}=77\pm5$~\kms. This
solution is shown with the folded radial velocity curve in
Figure~\ref{mt605curve}. The solution also results in masses of
$M_{1}\mathrm{sin}^{3}i=1.3\pm0.3$ and
$M_{2}\mathrm{sin}^{3}i=0.7\pm0.1$, indicating a mass ratio of
$q=0.57\pm0.05$. This computed mass ratio suggests that the secondary
may be 2--3 subclasses later than the primary, and a main-sequence
spectral type of B2.5:V would be in better agreement with the observed
mass ratio than our original determination of a B1 from analysis of
the line profiles alone. In addition, a relatively low inclination of
$i\sim26\degr$ would be required for the component masses to be
consistent with the theoretical masses of a B0.5V and B2.5V
\citep{drilling}.

\begin{figure}
\centering
\epsscale{0.9}
\plotone{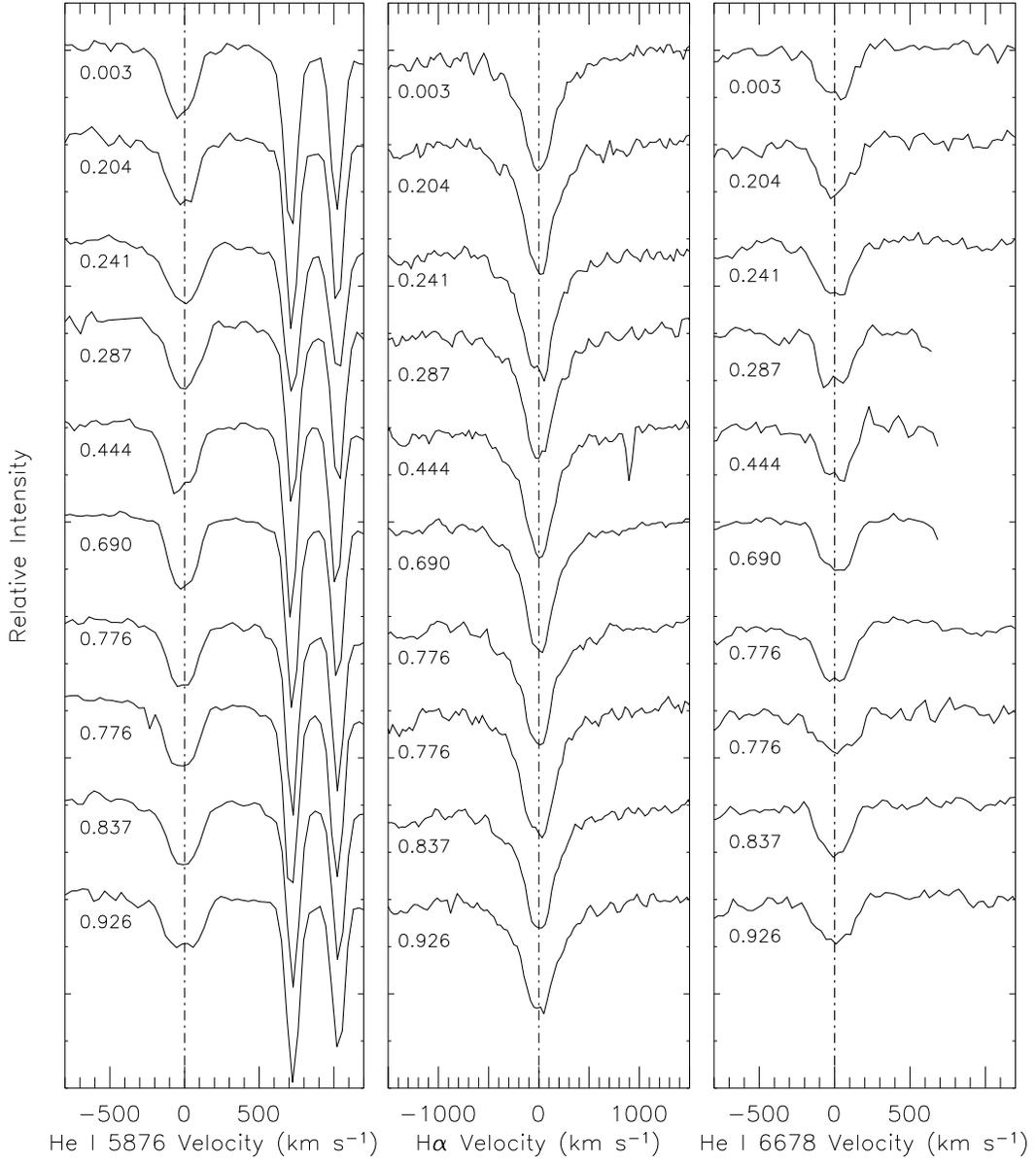}
\caption{\ion{He}{1}~$\lambda$5876~\AA\ (left),
H$\alpha$~$\lambda$6563~\AA\ (middle), \ion{He}{1}~$\lambda$6678~\AA\
(right) in velocity space and in order of phase for observations of
MT605 taken at WIRO between 2008 June 25 and 2009 August 4.
\label{mt605progress}}
\end{figure}

A time series of the 10 spectra used in the analysis, in order of
phase, is provided for illustrative purposes in
Figure~\ref{mt605progress}. Panel~1 shows the line profiles of
\ion{He}{1}~$\lambda$5876~\AA\, panel~2 shows the line profiles of
H$\alpha$, and panel~3 shows the line profiles of
\ion{He}{1}~$\lambda$6678~\AA.

\subsection{MT696}
\begin{figure}
\centering
\epsscale{0.9}
\plotone{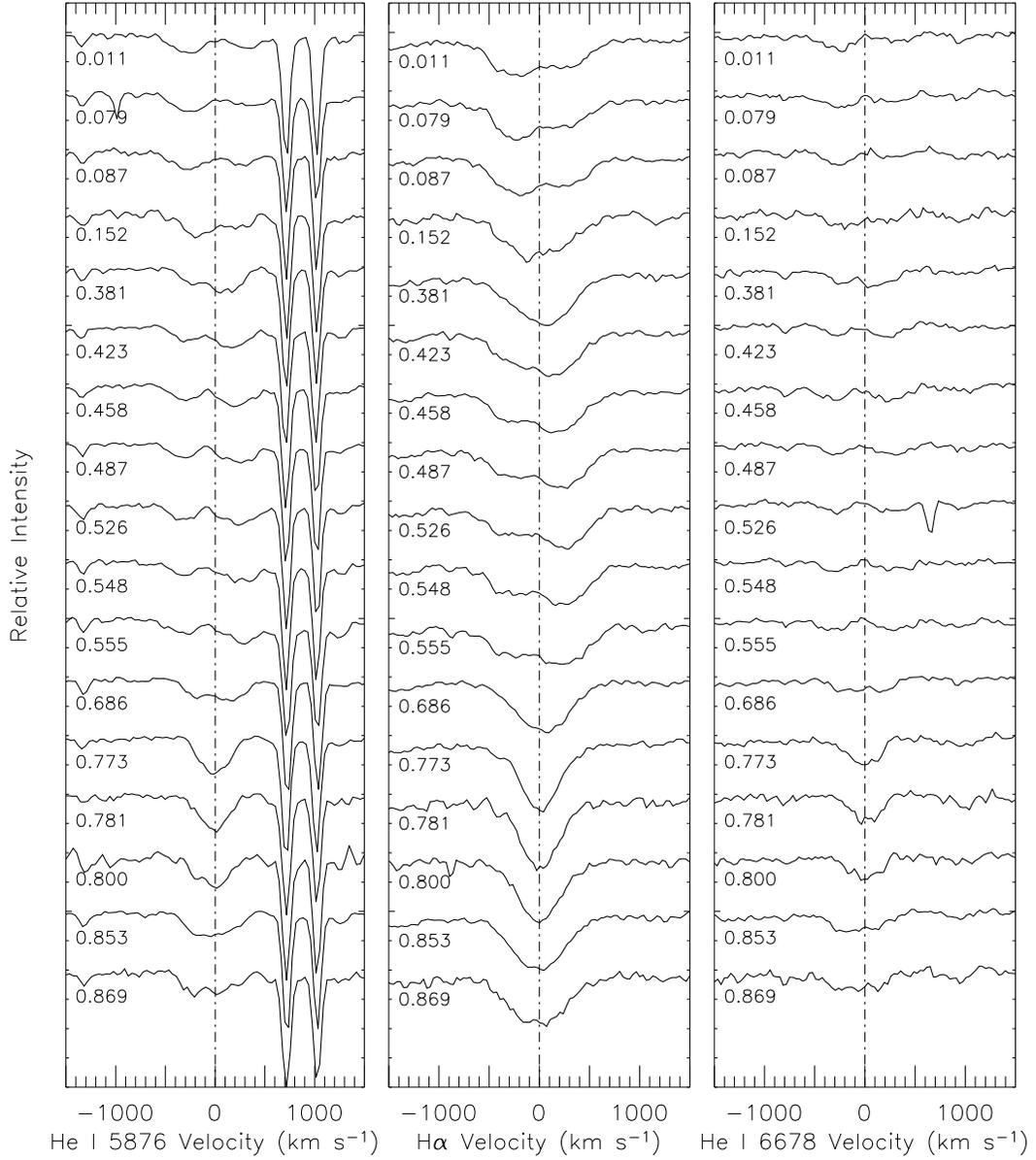}
\caption{\ion{He}{1}~$\lambda$5876~\AA\ (left),
H$\alpha$~$\lambda$6563~\AA\ (middle), \ion{He}{1}~$\lambda$6678~\AA\
(right) in velocity space and in order of phase for observations of
MT696.
\label{696progression}}
\end{figure}

\citet{Rios04} reveal the binary status of this interacting, eclipsing
system composed of an O9.5V and an early B star. The authors find a
period of $P=1.46$~days and present spectra also showing the SB2
nature of the system. We present the first spectroscopic solution for
this binary using 17 spectra obtained with WIRO--Longslit from 2008
August 20 through 2008 September 19. Owing to the large profile widths
of the \ion{He}{1}~$\lambda$5876~\AA\ lines and a maximum
line-velocity separation of $\sim$550~\kms, the positively shifted
\ion{He}{1}~$\lambda$5876~\AA\ profiles and the interstellar
\ion{Na}{1}~$\lambda$5890~\AA\ line are sometimes partially
blended. The broad spectral features, present in both components and
possibly indicative of large rotational velocities, is shown in
Figure~\ref{696progression}. The figure presents a time progression of
the \ion{He}{1}~$\lambda$5876~\AA, H$\alpha$, and $\lambda$6678~\AA\
lines (left, middle, and right panels respectively) in order of
phase. \citet{Rios04} report this system as W~UMa type, but the spectra
show no evidence of emission lines which are common in contact
binaries and systems with accretion disks. Given that this system has
such a short period, the possibility that this is (or has been) a
contact binary cannot be ruled out, but the early spectral types of
the components, the current absence of emission features, and the
longer than 1~day period suggest that this may in fact be a Beta~Lyr
type system instead.

\begin{figure}
\centering
\epsscale{0.9}
\plotone{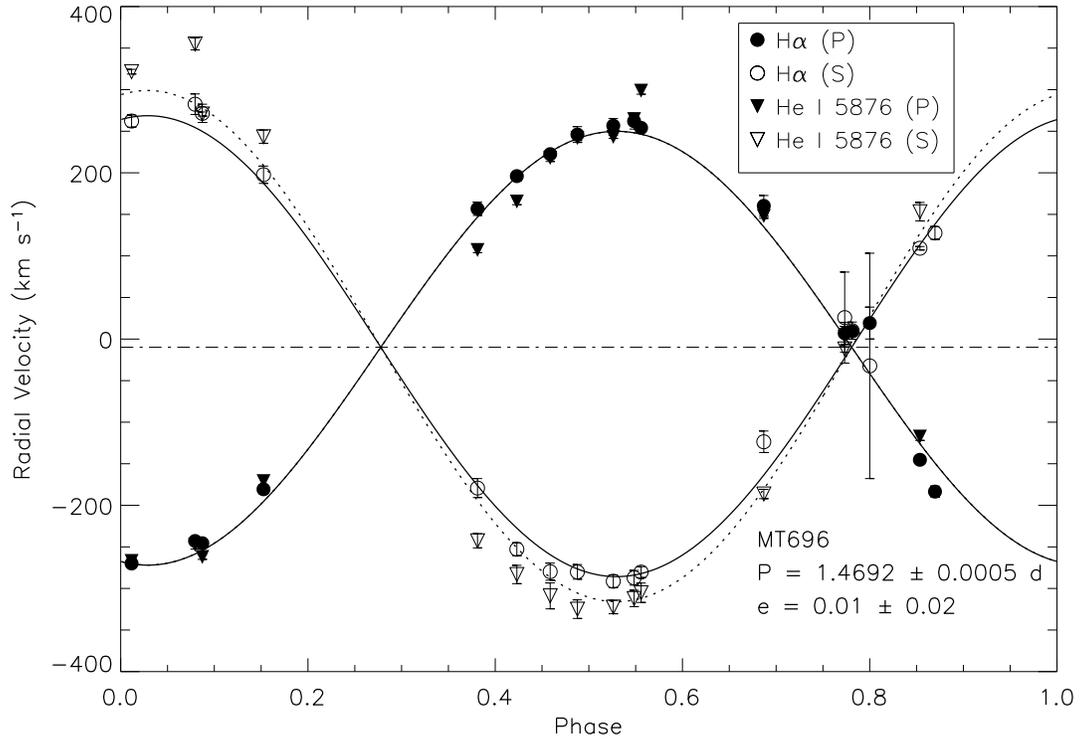}
\caption{Heliocentric radial velocity curves and orbital solutions for the
eclipsing binary (W~Uma type), MT696 (O9.5V + B0V) using 17 spectra
obtained at WIRO with WIRO-Longslit. The circles and solid line
represent the velocities and orbital solution obtained from H$\alpha$
and the triangles and dotted line represent the velocities and orbital
solution obtained from \ion{He}{1}~$\lambda$5876~\AA\. Filled and open
symbols correspond to the primary and secondary respectively.
\label{696curve}}
\end{figure}

We computed orbital solutions based on both the
\ion{He}{1}~$\lambda$5876~\AA\ velocities and H$\alpha$ velocities.
The \ion{He}{1} and H$\alpha$ spectroscopic periods for both
components agreed with \citet{Rios04} and showed strong signals in the
power spectra at $\sim$1.46~days. The orbital solution based on the
H$\alpha$ velocities, illustrated by the solid line in
Figure~\ref{696curve}, yields a period of $P=1.4692\pm0.0005$~days,
eccentricity of $e=0.01\pm0.02$, semi-amplitudes of
$K_{1}=261\pm4$~\kms\ and $K_{2}=277\pm6$~\kms, and mass ratio of
$q=0.94\pm0.02$. The H$\alpha$ radial velocities are represented by
the open (secondary) and filled (primary) circles, and the
\ion{He}{1}~$\lambda$5876~\AA\ velocities are represented by open and
filled triangles. The orbital solution based on the
\ion{He}{1}~$\lambda$5876~\AA\ velocities yields a period of
$P=1.467\pm0.004$~days, eccentricity of $e=0.05\pm0.02$,
semi-amplitudes of $K_{1}=261\pm6$~\kms\ and $K_{2}=327\pm10$~\kms,
and mass ratio of $q=0.80\pm0.03$. The period, eccentricity, epoch of
periastron, longitude of periastron, and systemic velocities for both
solutions showed good agreement.  We find a systemic velocity of
$\gamma=-10\pm3$, consistent with other Cyg~OB2 members. The secondary
semi-amplitudes, however, differ by $\sim$50~\kms. The more positive
semi-amplitude of the H$\alpha$ line may be a result of some
interaction between the two stars as H$\alpha$ is more easily
distorted in such scenarios because of its formation higher in the
stellar atmosphere or possibly in an extended photosphere/envelope.
However, given that the H$\alpha$ line profiles are high S/N, we
adopted the period, eccentricity, systemic velocity, angle of
periastron, and epoch of periastron from this solution and fixed them
in the determination of a revised \ion{He}{1}~$\lambda$5876~\AA\
solution. This is represented by the dotted line in
Figure~\ref{696curve}. The ephemerides for both the H$\alpha$ solution
and the combined H$\alpha+$\ion{He}{1}~$\lambda$5876~\AA\ solution are
provided in Table~\ref{OC24}. The orbital elements to the combined
H$\alpha+$\ion{He}{1}~$\lambda$5876~\AA\ solution are provided in
Table~\ref{orbparms4.tab}.

Adopting the revised \ion{He}{1} orbital solution, we obtain a mass
ratio of $0.85\pm0.03$.  This, coupled with the nearly equal line
depths for all features, suggests a secondary spectral type similar
to the primary's. Without the photometric data, we can only speculate
on the inclination of MT696. However, assuming a theoretical mass of
16$\pm$0.5~\msun\ for the primary \citep{FM05}, the calculated masses
($M_{1}\mathrm{sin}^{3}i=15.1\pm0.7$~\msun\ and
$M_{2}\mathrm{sin}^{3}i=12.8\pm0.5$~\msun) suggest that the system has
an inclination of near 80$^\circ$ and a secondary with a spectral type
of B1V.

\subsection{MT720}
\begin{figure}
\centering
\epsscale{0.9}
\plotone{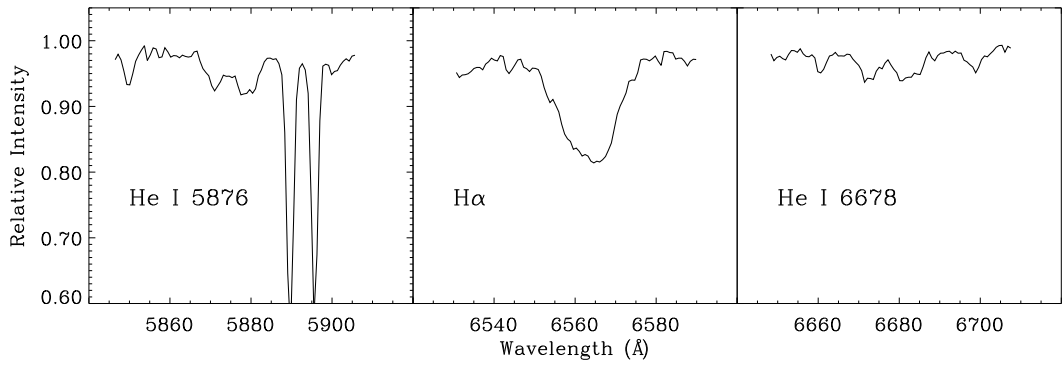}
\caption{A composite of seven spectra between a phase of $\phi=0.5$
and $\phi=0.7$ for MT720. The three panels show \ion{He}{1}~$\lambda$5876~\AA,
H$\alpha$, and \ion{He}{1}~$\lambda$6678~\AA\ at quadrature respectively.
\label{720composite}}
\end{figure}

In Paper~II, we find that the SB2 MT720 is most likely composed of two
early B stars based on the absence of \ion{Mg}{2}~$\lambda$4481~\AA\ and
strength of the \ion{He}{1} lines. Based on new WIRO-Longslit
observations of this system, we agree with our previous
assessment. The new spectra show relatively strong \ion{He}{1}
absorption and no evidence of \ion{He}{2}. The paucity of stellar
lines in this spectral regime and the low S/N of our spectra still
make accurate spectral typing difficult. Figure~\ref{720composite}
shows the dominant stellar lines in the 5250--6700~\AA\ regime
(\ion{He}{1}~$\lambda$5876~\AA, H$\alpha$, and
\ion{He}{1}~$\lambda$6678~\AA) near quadrature. The spectrum is a
composite of seven spectra between $\phi=0.5$ and $\phi=0.7$. The line
depths and widths are similar and suggest comparable spectral types for
the components. In order to draw further conclusions regarding the
component spectral types, we examined the following orbital solution
and component mass estimates.

\begin{figure}
\centering
\epsscale{0.9}
\plotone{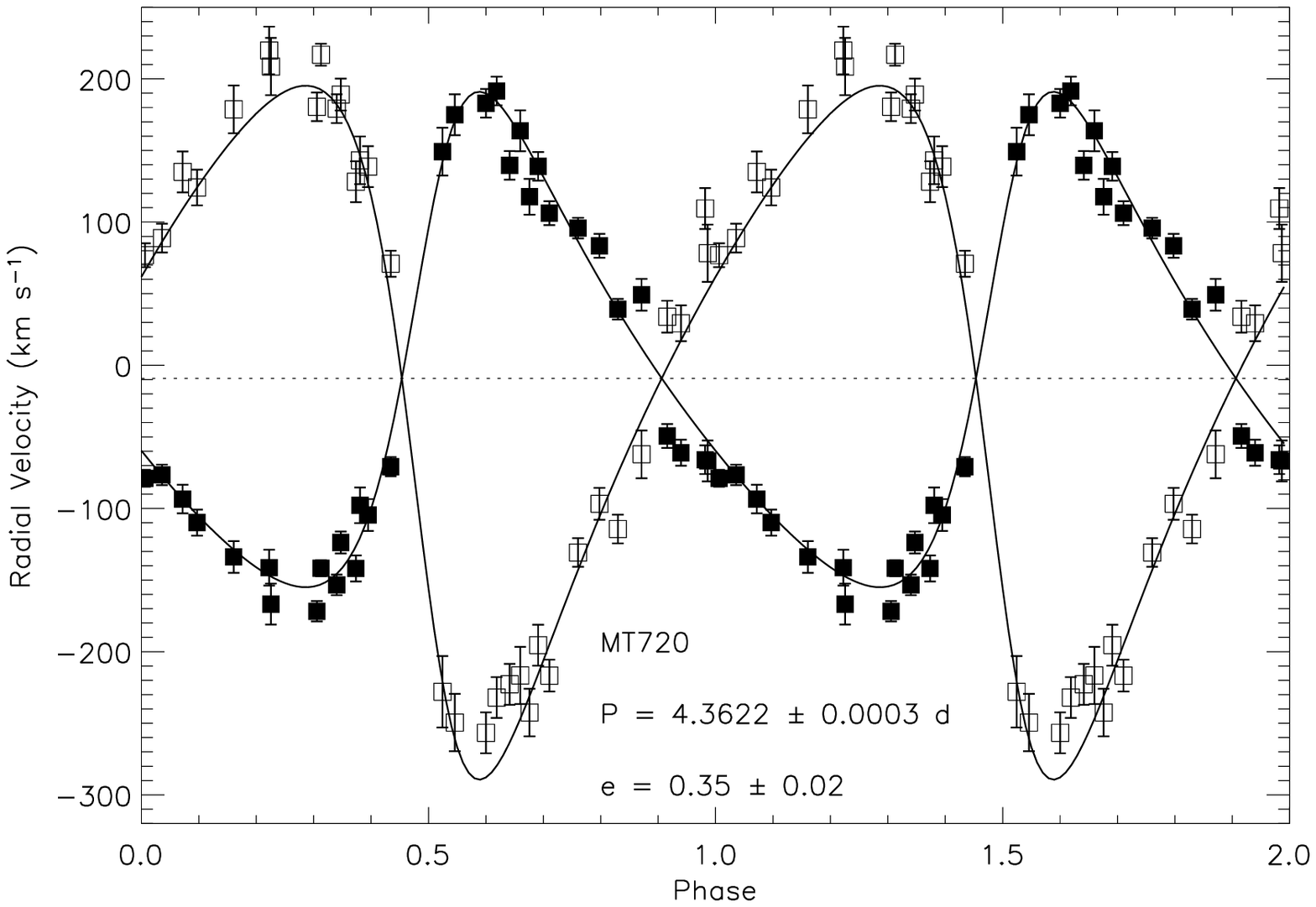}
\caption{Heliocentric radial velocity curve and orbital solution for
MT720 using 32 of the highest S/N spectra obtained at WIRO with
WIRO-Longslit.
\label{720curve}}
\end{figure}

We measured radial velocities for 32 of the 35 WIRO-Longslit spectra
obtained on MT720 using the \ion{He}{1}~$\lambda$5876~\AA\
profiles. These 32 radial velocities provide a clear period of
4.36~days from the power spectra of both components. The power spectra
also contained no significant aliases. The 4.36~day period agrees with
our original visual assessment of just under 5~days in Paper~II. The
best combined orbital solution obtained from BSCSP produces a period
of $P=4.3622\pm0.0003$~days and an eccentricity of $e=0.35\pm0.02$.
Figure~\ref{720curve} displays this solution. We note that this
solution presents an unusually large eccentricity for such a
short-period system. We computed additional solutions with an
eccentricity fixed at lower values, but these resulted in a much
larger rms ($>>$21.6 \kms\ for the primary and $>>$29.5 \kms\ for the
secondary). Given semi-amplitudes of $K_{1}=173\pm5$~\kms\ and
$K_{2}=242\pm7$~\kms, the calculated mass ratio is $q=0.71\pm0.06$.
Additionally, the calculated minimum masses,
$M_{1}\mathrm{sin}^{3}i=15.5 \pm 0.8$~\msun\ and
$M_{2}\mathrm{sin}^{3}i=11.1 \pm 0.6$~\msun, indicate that the likely
spectral types of the components are B0--1V and B1--2V for the primary
and secondary respectively \citep[interpolated
from][]{drilling}. There is no record of photometric variability for
this system, but if the primary has a B0V or later spectral type, the
minimum masses of the components preclude inclinations lower than
$i\sim70$\deg. The remaining orbital parameter values are listed in
Table~\ref{orbparms4.tab}.

\subsection{MT771}
\begin{figure}
\centering
\epsscale{0.9}
\plotone{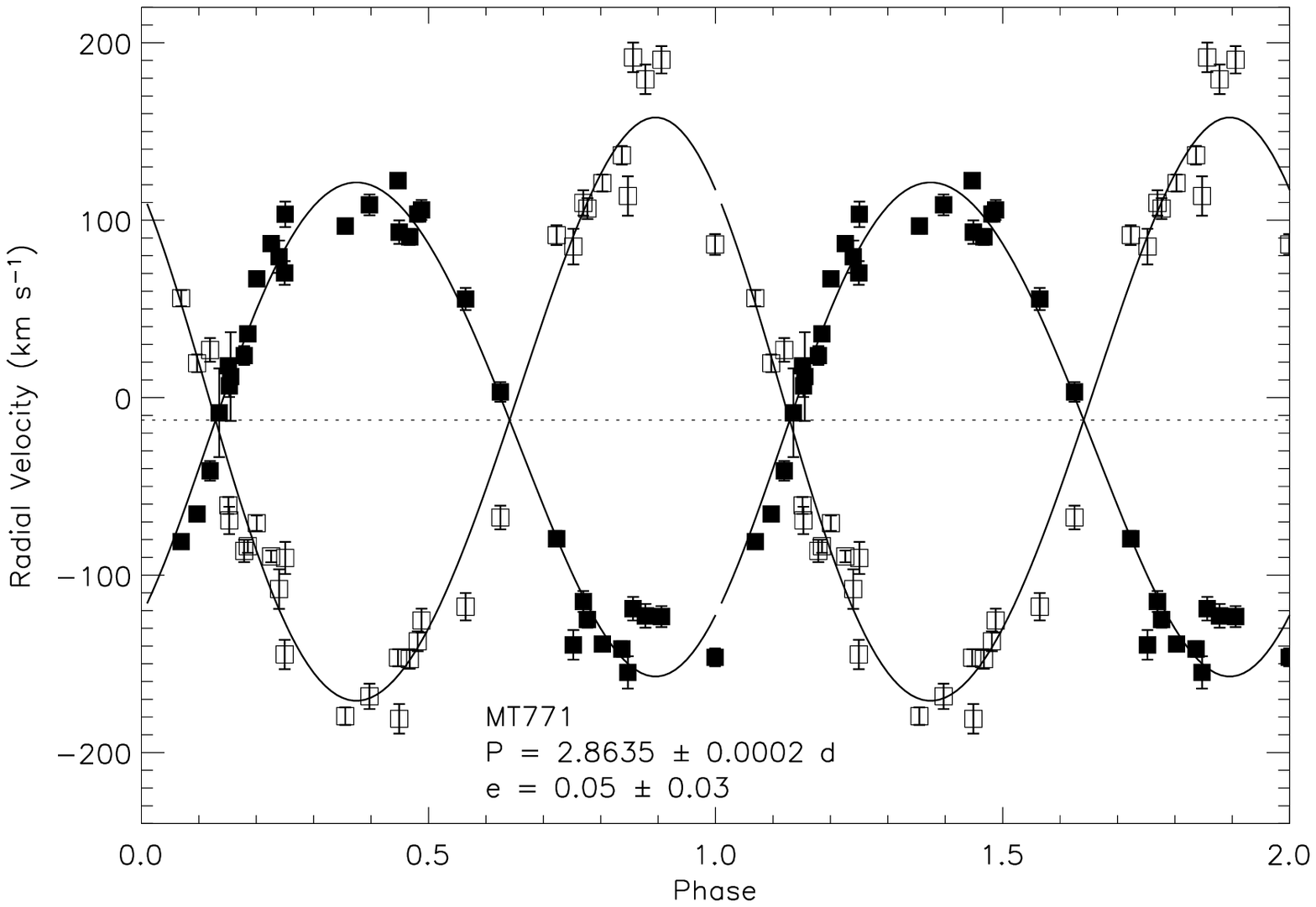}
\caption{Heliocentric radial velocity curve and orbital solution for
MT771 using 34 of the highest S/N spectra obtained at WIRO with
WIRO-Longslit.\label{771curve}}
\end{figure}

\begin{figure}
\centering
\epsscale{0.9}
\plotone{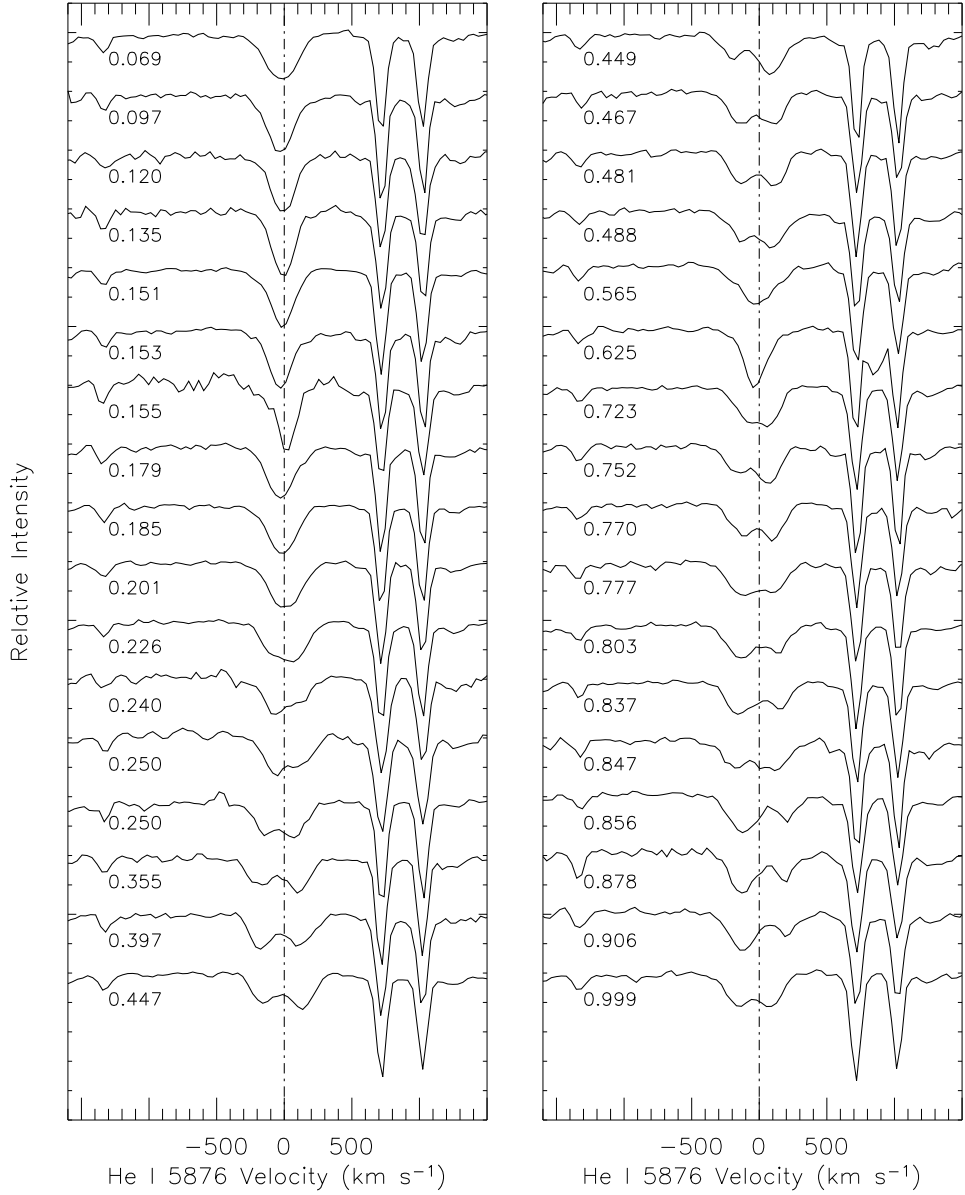}
\caption{\ion{He}{1}~$\lambda$5876~\AA\ in velocity space and in order of
phase for observations of MT771 obtained at WIRO with
WIRO-Longslit from 2007--2009.
\label{plot771}}
\end{figure}

The \ion{He}{1} lines of the SB2, late O-type binary MT771 show a
strong Struve-Sahade effect \citep[for a good summary of this effect,
see][]{Linder07}.  This effect causes asymmetries or varying line
depth/width and, in the case of MT771, is present in both components'
spectra. This makes component identification tricky for such a short
period ($<$ 3~days) system, even with a maximum velocity separation of
$\sim$300~\kms. The much broader H$\alpha$ line remains heavily
blended even at the largest velocity separation. We used this to our
advantage, following the same procedure as with MT605, fitting a
single Gaussian profile to the H$\alpha$ line and measuring the
Gaussian FWHM. The period produced by the FWHM measurements is equal
to half the binary period. The FWHM measurements yielded a strong
signal in the power spectrum corresponding to a period of 1.43~days
with no significant aliases. Even with the varying asymmetries in the
component \ion{He}{1} profiles, we still chose to use the average
measured Gaussian width and depth for each component's
\ion{He}{1}~$\lambda$5876~\AA\ line and fix these values when
measuring the radial velocities (assuming a period of
$\sim$2.86~days). Of the 34 epochs included for MT771 in
Table~\ref{OC24}, we omitted two observations for the secondary ($HJD
= 2,454,700.85$ \&\ $2,455,041.67$) when the lines were at their most
blended state. As was expected, the power spectra for both components
showed a clear, strong signal at 2.86~days. The computed orbital
solution for this system produces a period of
$P=2.8635\pm0.0002$~days, an eccentricity of $e=0.05\pm0.03$, and
semi-amplitudes of $K_{1}=139\pm5$~\kms\ and $K_{2}=164\pm6$~\kms. The
solution also produces lower limit component masses of
$M_{1}\mathrm{sin}^{3}i=4.5\pm0.3$~\msun\ and $M_{2}\mathrm{sin}^{3}i
=3.8\pm0.2$~\msun\ and suggests a low inclination of $i\lesssim
37^\circ$ for this system \citep[the approximate inclination required
to yield the mass of a late O star;][]{FM05}. This solution and the
radial velocity curve, produced from the 34 WIRO-Longslit spectra
obtained between 2007 August 28 and 2009 August 3, is shown in
Figure~\ref{771curve}.  The radial velocities measured at
$\phi=$0.8--1.0 appear systematically more positive, and this likely
stems from the strong asymmetries at these epochs.  This is more
easily seen in Figure~\ref{plot771}, which shows a time series of the
\ion{He}{1}~$\lambda$5876~\AA\ line in order of phase. Panel 1
contains the line profiles for $\phi=$0.07--0.45 and panel 2 contains
the line profiles for $\phi=$0.45--1.00. The
\ion{He}{1}~$\lambda$5876~\AA\ line of the primary appears much
stronger than the secondary at phases of $\phi=$0.856--0.906, and the
secondary also appears asymmetric. The ratio of He line depth also
appears to change over a small range of phase (e.g.,
$\phi=$0.355--0.387), though less dramatically.

The strong Struve-Sahade effect indicates that there are likely
significant irradiation effects in this system,
possibly stemming from a shock region between the components' stellar
winds. If so, the shock region is nearer the secondary as the
asymmetry is stronger in the secondary's spectral lines. The asymmetry
may be the result of an unresolved emission feature between the
component lines and originate from the shock region. The seemingly
irregular strength of the emission would indicate varying wind
velocities and densities of one or both stars. 

The Struve-Sahade effect also makes spectral typing difficult given
that the \ion{He}{1} and \ion{He}{2} lines are crucial to temperature
classification in late O stars (e.g.,
\ion{He}{2}~$\lambda$4541~\AA:\ion{He}{1}~$\lambda$4471~\AA,
\ion{He}{2}~$\lambda$4541~\AA:\ion{Si}{3}~$\lambda$4552~\AA, and
\ion{He}{1}~$\lambda$4471 \AA:\ion{Mg}{2}~$\lambda$4481 \AA). Our
initial classifications of O7V for the primary and O9V for the
secondary, however, appear to still be close given the new computed
mass ratio of $q=0.85\pm0.04$, the moderately strong \ion{He}{2} at
$\lambda\lambda$4200, 4542, and 5411~\AA, and strong \ion{He}{1}.

\section{Summary}
We present orbital solutions for two new SB2s, MT311 and MT605. We
also provide the first spectroscopic solutions to the eclipsing
system, MT696, the early B, eclipsing triple SB1, MT429, the early B
SB2, MT720, and the late-O SB2, MT771.

\begin{figure}
\centering
\epsscale{0.9}
\plotone{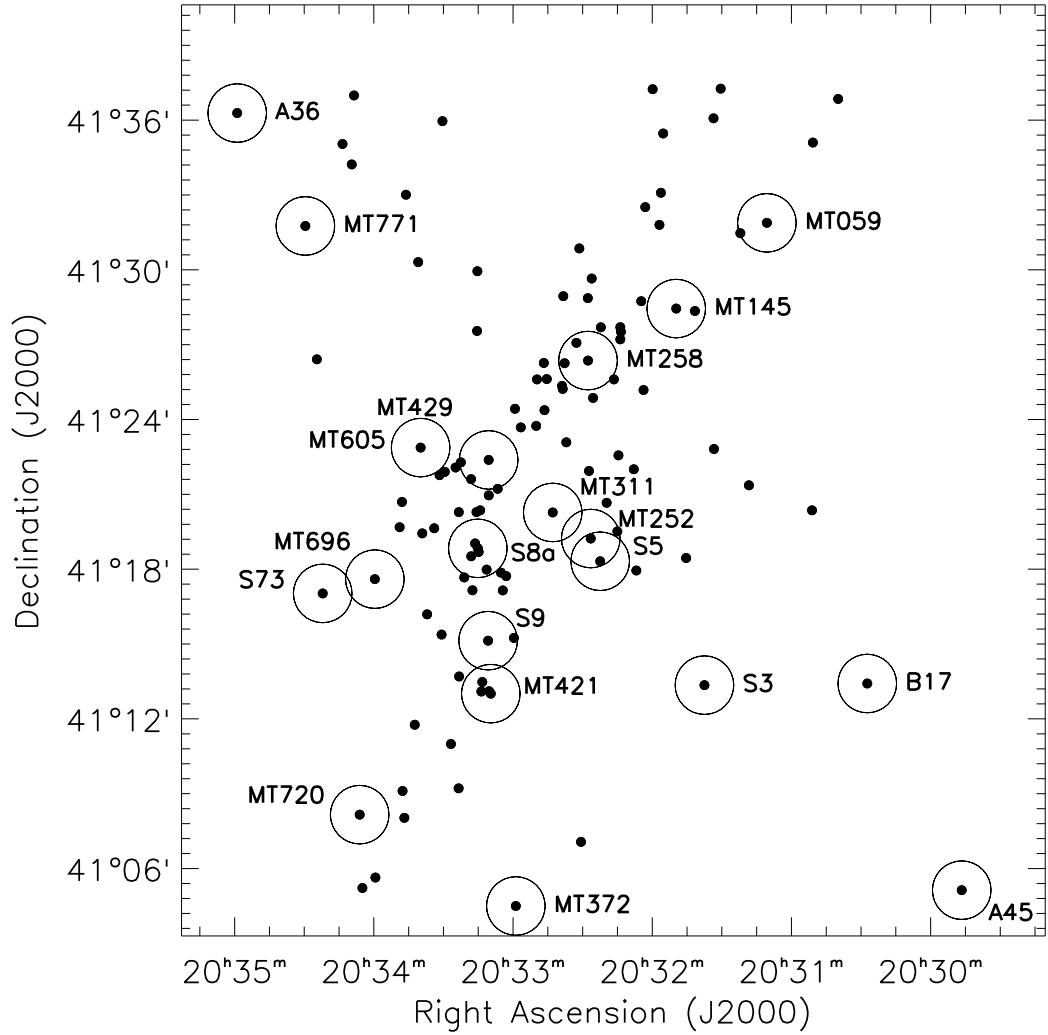}
\caption{Map of OB stars in Cyg~OB2. Circles indicate the 
location of the 20 known binaries within Cyg~OB2. We find no apparent
clusterings or groups of massive stars, nor a tendency for them to 
located near the center of the Association.
\label{binaries4}}
\end{figure}

Table~\ref{Binaries4} lists the 20 known OB binaries in
Cyg~OB2. Column~1 gives the system designations.  Column~2 shows the
binary types (i.e., eclipsing, SB1, SB2, etc). Column~3 provides the
spectral types of the components. The spectral types in square
brackets have been determined indirectly from the data (i.e., based on
the computed mass ratio rather than the component's
spectrum). Columns~4--6 give the orbital periods, eccentricities, and
mass-ratios respectively. Finally, column~7 lists the corresponding
surveys responsible for the orbital
information. Figure~\ref{binaries4} shows the locations of these
systems relative to the other OB stars in this survey.  There are no
apparent clusterings or groups of massive stars, nor are there
tendencies of the massive binaries to reside in the center of the
Association versus the outskirts.  Of the 20, we know reliably the
orbital periods of all 20, the orbital eccentricities of 18, and the
mass-ratios of all 20 (the mass ratios for the SB1s are actually range
estimates based on the range of probable inclinations). There are six
SB1s, 14 SB2s, five eclipsing binaries of the Algol type, two
eclipsing binaries of the $\beta$~Lyr type, and one eclipsing binary
of the W~UMa type. At least five binaries are both spectroscopic and
eclipsing systems. In a companion paper \citep{Kiminki11b}, we use the
orbital parameter information and ephemerides associated with the
short period systems ($P<26$~days) listed in Table~\ref{Binaries4},
along with the ephemerides of 94 additional OB stars in Cyg~OB2, to
compute the intrinsic binary orbital parameter distributions (i.e.,
period, mass ratio, and eccentricity) and intrinsic binary fraction of
the cluster.

\acknowledgements We would like to acknowledgment the time, effort,
and thoroughness of our anonymous referee in assisting to make this a
much stronger manuscript. We thank the time allocation committees of
the Lick, Keck, WIYN, and WIRO observatories for granting us observing
time and making this project possible. We are also appreciative to
Chris Fryer, Falk Herwig, and Jean-Claude Passy for providing useful
discussion on the intriguing system MT429, and Saida Caballero-Nieves
for providing a big piece to the MT429 puzzle in the form of her HST
data. Additionally, I would like to thank Steward Observatory at the
University of Arizona for providing support and resources during my
transition from the University of Wyoming and the final stages of this
manuscript. And finally, we are thankful for the continued support
from the National Science Foundation through Research Experience for
Undergraduates (REU) program grant AST 03-53760, through grant AST
03-07778, and through grant AST 09-08239, and the support of the
Wyoming NASA Space Grant Consortium through grant NNG05G165H.
\textit{Facilities:} \facility{WIRO ()}, \facility{WIYN ()},
\facility{Shane ()}, \facility{Keck:I ()}

\thispagestyle{empty}

\clearpage

\begin{deluxetable}{lc}
\centering
\tabletypesize{\scriptsize}
\tablecaption{Dates of WIRO Observing Runs \label{obs4.tab}}
\tablewidth{0pt}
\tablehead{
\colhead{} &
\colhead{Date Coverage} \\
\colhead{Date} &
\colhead{(HJD)}}
\startdata
2008 Dec 17                & 2,454,818 \\
2009 Jan 3                 & 2,454,835 \\
2009 Jan 16                & 2,454,848 \\
2009 Jun 27--30            & 2,455,010--2,455,013 \\
2009 Jul 5                 & 2,455,018 \\ 
2009 Jul 27,28             & 2,455,040--2,455,041 \\
2009 Jul 30--Aug 3         & 2,455,043--2,455,047 \\
2009 Aug 26--29            & 2,455,070--2,455,073 \\
2009 Aug 31--Sep 4         & 2,455,075--2,455,079 \\
2010 Aug 21                & 2,455,430 \\
2010 Oct 27                & 2,455,497 \\
\enddata
\end{deluxetable}

\clearpage

\begin{deluxetable}{lcrrrr}
\centering
\tabletypesize{\tiny}
\tabletypesize{\scriptsize}
\tablewidth{0pc}
\tablecaption{Ephemerides for MT311, MT696, MT720, \&\ MT771 \label{OC24}}
\tablehead{
\colhead{} & 
\colhead{} & 
\colhead{$V_{r1}$} &
\colhead{$O_1-C_1$} &
\colhead{$V_{r2}$} &
\colhead{$O_2-C_2$} \\ 
\colhead{Date (HJD-2,400,000)} &
\colhead{$\phi$} &
\colhead{(\kms)} &
\colhead{(\kms)} &
\colhead{(\kms)} &
\colhead{(\kms)}}
\tablecolumns{6}
\startdata
\cutinheadc{MT311}
52,146.68..........................  &   0.569    &    -98.0 (10.5) &   -13.2  &    104.5 (22.4) &  -11.9  \\
55,010.64..........................  &   0.457    &    -87.1 (6.2)  &   -10.4  &    101.1 (11.0) &   -4.4  \\
55,011.78..........................  &   0.655    &    -75.7 (6.8)  &   -12.8  &     83.8 (12.0) &   -3.3  \\
55,012.72..........................  &   0.818    &     27.2 (6.3)  &     7.6  &    -62.3 (11.0) &  -39.2  \\
55,013.82..........................  &   0.010    &     77.8 (4.9)  &   -10.8  &   -104.1 (8.5)  &   11.1  \\
55,070.66..........................  &   0.891    &     67.5 (6.8)  &    11.4  &   -123.3 (11.9) &  -51.5  \\
55,071.72..........................  &   0.075    &     72.7 (4.7)  &   -14.2  &    -91.6 (8.4)  &   21.2  \\
55,076.81..........................  &   0.961    &     81.6 (6.9)  &     1.3  &    -91.6 (12.2) &   12.5  \\
55,077.74..........................  &   0.122    &     86.5 (5.1)  &     9.9  &   -104.1 (9.0)  &   -4.9  \\
55,078.73..........................  &   0.293    &     20.8 (5.1)  &    24.3  &     \nodata     & \nodata \\
55,079.76..........................  &   0.472    &    -81.6 (5.1)  &    -1.4  &    132.3 (9.0)  &   21.9  \\
\cutinhead{MT605}
54,643.87..........................  &   0.241     &    -52.8 (4.9)   &    -3.7  &    65.8 (6.8)  &   11.8 \\
54,674.94..........................  &   0.776     &     22.2 (5.1)   &    -0.8  &   -64.7 (7.2)  &    4.4 \\
55,043.65..........................  &   0.837     &     26.6 (5.9)   &    -6.3  &   -75.9 (8.2)  &   13.3 \\
55,044.75..........................  &   0.926     &     42.6 (7.2)   &     5.4  &  -107.8 (10.0) &  -11.1 \\
55,045.68..........................  &   0.003     &      7.5 (6.3)   &    -9.7  &   -58.0 (8.8)  &    4.0 \\
55,072.64..........................  &   0.204     &    -47.7 (7.5)   &    -1.4  &    38.4 (10.4) &  -10.6 \\
55,073.69..........................  &   0.287     &    -36.3 (7.4)   &    13.3  &    49.8 (10.4) &   -4.9 \\
55,075.62..........................  &   0.444     &    -48.0 (6.1)   &   -10.6  &    37.8 (8.4)  &    4.3 \\
55,078.63..........................  &   0.690     &      7.1 (8.7)   &     3.5  &   -31.0 (12.2) &    6.9 \\
55,079.68..........................  &   0.776     &     27.4 (7.7)   &     6.0  &   -77.4 (10.7) &   -8.3 \\
\cutinhead{MT696 (\ion{He}{1}~$\lambda$5876~\AA~$+$~H$\alpha$)}
54,699.74..........................  & 0.080  & -247.5 (5.2)    &    9.5   &   355.0 (7.2)    &   73.7 \\
54,701.91..........................  & 0.556  &  298.8 (4.2)    &   50.3   &  -305.2 (11.7)   &    9.0 \\
54,724.62..........................  & 0.012  & -266.9 (3.7)    &    1.0   &   321.6 (2.8)    &   27.4 \\
54,724.73..........................  & 0.087  & -262.1 (2.9)    &   -9.4   &   270.7 (2.4)    &   -5.7 \\
54,724.82..........................  & 0.153  & -170.9 (4.9)    &   25.3   &   243.4 (7.9)    &   33.6 \\
54,725.61..........................  & 0.687  &  147.9 (2.7)    &   15.6   &  -186.4 (5.6)    &   -9.1 \\
54,725.74..........................  & 0.773  &    2.8 (12.5)   &    4.4   &   -11.9 (4.0)    &    7.6 \\
54,725.85..........................  & 0.854  & -117.4 (4.5)    &   10.8   &   153.2 (11.1)   &   23.6 \\
54,726.63..........................  & 0.381  &  107.2 (3.3)    &  -39.0   &  -242.6 (8.5)    &  -49.1 \\
54,726.74..........................  & 0.459  &  217.9 (4.3)    &   -9.1   &  -308.9 (15.6)   &  -20.1 \\
54,726.84..........................  & 0.526  &  243.2 (1.9)    &   -9.0   &  -322.4 (8.0)    &   -3.8 \\
54,729.63..........................  & 0.423  &  165.4 (3.8)    &  -30.5   &  -283.2 (11.0)   &  -31.0 \\
54,729.72..........................  & 0.488  &  241.8 (3.0)    &   -1.7   &  -324.9 (11.2)   &  -16.6 \\
54,729.81..........................  & 0.548  &  264.5 (4.5)    &   14.3   &  -312.1 (9.8)    &    4.2 \\
\cutinheadb{MT696 (H$\alpha$)}
54,699.74..........................  & 0.079  & -242.9 (4.7)    &   14.4  &  282.5 (12.3)     &  29.4  \\
54,700.77..........................  & 0.781  &    9.3 (6.0)    &   22.6  &   10.1 (10.4)     &  16.3  \\
54,700.90..........................  & 0.869  & -183.5 (6.6)    &  -33.7  &  127.6 (7.8)      & -11.2  \\
54,701.91..........................  & 0.555  &  254.1 (4.5)    &    5.4  & -280.3 (6.6)      &   4.2  \\
54,722.84..........................  & 0.800  &   19.2 (19.2)   &   63.2  &  -32.3 (135.5)    & -58.7  \\
54,724.62..........................  & 0.011  & -269.9 (5.0)    &   -2.1  &  262.2 (7.3)      &  -2.0  \\
54,724.73..........................  & 0.087  & -245.6 (4.8)    &    7.6  &  271.5 (11.0)     &  22.7  \\
54,724.82..........................  & 0.152  & -180.7 (5.5)    &   16.5  &  197.7 (10.4)     &   8.5  \\
54,725.61..........................  & 0.686  &  160.3 (12.5)   &   26.8  & -123.6 (12.9)     &  38.5  \\
54,725.74..........................  & 0.773  &    7.0 (13.0)   &    7.3  &   25.8 (54.8)     &  45.8  \\
54,725.85..........................  & 0.853  & -145.2 (4.8)    &  -18.2  &  109.3 (2.1)      &  -5.4  \\
54,726.63..........................  & 0.381  &  156.7 (8.0)    &   11.6  & -179.2 (11.4)     &  -4.8  \\
54,726.74..........................  & 0.458  &  222.5 (4.4)    &   -3.9  & -279.8 (10.3)     & -19.0  \\
54,726.84..........................  & 0.526  &  256.7 (8.7)    &    4.5  & -291.3 (7.7)      &  -3.0  \\
54,729.63..........................  & 0.423  &  196.0 (5.8)    &    1.0  & -252.8 (7.8)      & -25.3  \\
54,729.72..........................  & 0.487  &  246.0 (9.5)    &    2.8  & -280.1 (8.8)      &  -1.4  \\
54,729.81..........................  & 0.548  &  262.1 (9.5)    &   11.7  & -287.2 (8.8)      &  -0.8  \\
\cutinhead{MT720}
54,341.83..........................  &  0.072   &   -93.4 (10.5)  &     0.1  &   135.0 (14.8) &   25.9 \\
54,343.81..........................  &  0.524   &   149.2 (15.9)  &     6.9  &  -228.0 (22.4) &   -7.0 \\
54,345.82..........................  &  0.986   &   -66.8 (14.4)  &   -14.0  &    78.2 (20.3) &   26.2 \\
54,346.87..........................  &  0.226   &  -166.8 (15.4)  &   -19.3  &   208.6 (21.7) &   23.8 \\
54,348.76..........................  &  0.660   &   163.7 (13.5)  &     3.0  &  -216.6 (19.0) &   30.4 \\
54,399.56..........................  &  0.306   &  -171.8 (7.1)   &   -17.8  &   180.5 (9.9)  &  -13.3 \\
54,399.71..........................  &  0.340   &  -153.4 (7.2)   &    -7.8  &   179.2 (10.1) &   -2.9 \\
54,401.55..........................  &  0.760   &    95.8 (7.3)   &     9.4  &  -130.8 (10.3) &   12.2 \\
54,401.71..........................  &  0.798   &    83.4 (8.3)   &    23.5  &   -96.8 (11.7) &    8.9 \\
54,402.62..........................  &  0.006   &   -79.0 (5.8)   &   -16.2  &    76.8 (8.2)  &   10.8 \\
54,402.75..........................  &  0.036   &   -76.5 (7.0)   &     0.6  &    88.8 (9.8)  &    2.7 \\
54,405.60..........................  &  0.691   &   138.9 (9.7)   &     0.5  &  -195.5 (13.6) &   20.2 \\
54,406.59..........................  &  0.916   &   -49.4 (7.8)   &   -34.7  &    33.9 (11.2) &   35.3 \\
54,408.58..........................  &  0.374   &  -141.9 (9.5)   &   -15.5  &   128.2 (13.3) &  -26.9 \\
54,409.57..........................  &  0.600   &   183.0 (9.9)   &    -7.0  &  -256.7 (13.9) &   31.2 \\
54,409.75..........................  &  0.641   &   139.6 (9.7)   &   -32.9  &  -222.8 (13.7) &   40.7 \\
54,410.57..........................  &  0.830   &    39.2 (7.2)   &     0.9  &  -114.3 (10.1) &  -39.0 \\
54,410.75..........................  &  0.871   &    49.2 (11.5)  &    37.3  &   -62.2 (16.1) &  -23.6 \\
54,643.88..........................  &  0.313   &  -141.7 (5.6)   &    11.3  &   216.9 (8.0)  &   24.4 \\
54,696.75..........................  &  0.434   &   -70.8 (6.5)   &   -23.0  &    70.9 (9.1)  &   25.8 \\
54,725.82..........................  &  0.097   &  -109.8 (8.8)   &    -5.5  &   124.1 (12.3) &   -0.1 \\
54,748.72..........................  &  0.347   &  -123.8 (7.8)   &    18.9  &   189.0 (11.0) &   11.1 \\
54,754.67..........................  &  0.710   &   106.2 (8.3)   &   -17.5  &  -216.7 (11.7) &  -21.6 \\
54,755.67..........................  &  0.940   &   -61.1 (8.9)   &   -33.1  &    29.3 (12.5) &   11.9 \\
54,757.65..........................  &  0.395   &  -104.6 (10.7)  &     1.7  &   138.7 (15.1) &   11.7 \\
55,046.79..........................  &  0.676   &   117.7 (12.9)  &   -31.7  &  -242.5 (18.2) &  -11.4 \\
55,070.71..........................  &  0.161   &  -133.8 (11.6)  &    -5.1  &   178.7 (16.3) &   20.2 \\
55,071.67..........................  &  0.381   &   -97.8 (11.9)  &    22.5  &   143.1 (16.7) &   -3.5 \\
55,072.71..........................  &  0.619   &   191.5 (10.5)  &     7.4  &  -232.0 (14.8) &   47.7 \\
55,076.76..........................  &  0.546   &   174.9 (13.8)  &     4.2  &  -249.5 (19.5) &   11.5 \\
55,078.66..........................  &  0.982   &   -66.0 (10.3)  &   -15.5  &   109.5 (14.4) &   60.6 \\
55,079.71..........................  &  0.223   &  -141.3 (12.4)  &     5.5  &   219.8 (17.4) &   36.0 \\
\cutinhead{MT771}
54,341.82..........................  &   0.752  &  -139.3 (8.1)  &   -38.9  &    85.1 (9.9)  &    -5.8 \\
54,342.81..........................  &   0.097  &   -65.6 (4.0)  &   -22.7  &    19.3 (4.9)  &    -3.7 \\
54,343.82..........................  &   0.449  &    93.3 (6.9)  &   -15.4  &  -181.0 (8.5)  &   -25.1 \\
54,345.83..........................  &   0.153  &     6.6 (6.3)  &    -2.7  &   -69.2 (7.7)  &   -30.6 \\
54,347.82..........................  &   0.847  &  -154.9 (8.8)  &    -5.0  &   113.7 (10.8) &   -35.7 \\
54,348.77..........................  &   0.179  &    23.7 (5.2)  &    -8.4  &   -86.4 (6.4)  &   -20.8 \\
54,397.65..........................  &   0.250  &    70.3 (6.5)  &   -13.4  &  -144.7 (8.1)  &   -18.3 \\
54,401.59..........................  &   0.625  &     3.3 (5.5)  &     2.9  &   -67.6 (6.8)  &   -39.5 \\
54,402.66..........................  &   0.999  &  -146.3 (4.9)  &   -23.2  &    86.3 (6.0)  &   -31.4 \\
54,403.68..........................  &   0.355  &    96.7 (4.1)  &   -23.7  &  -179.5 (5.0)  &    -9.7 \\
54,696.81..........................  &   0.723  &   -79.5 (4.4)  &    -0.4  &    91.6 (5.4)  &    25.9 \\
54,696.94..........................  &   0.770  &  -114.9 (5.7)  &    -2.7  &   109.7 (7.0)  &     4.9 \\
54,697.80..........................  &   0.069  &   -81.1 (3.7)  &   -12.9  &    56.0 (4.5)  &     3.1 \\
54,697.95..........................  &   0.120  &   -41.3 (5.5)  &   -19.6  &    26.9 (6.8)  &    28.9 \\
54,698.74..........................  &   0.397  &   108.6 (5.8)  &   -11.4  &  -168.3 (7.1)  &     1.0 \\
54,698.94..........................  &   0.467  &    90.6 (5.6)  &   -11.2  &  -147.3 (5.6)  &     0.5 \\
54,699.83..........................  &   0.777  &  -125.0 (4.6)  &    -8.3  &   106.6 (5.7)  &    -3.7 \\
54,700.85..........................  &   0.135  &    -8.6 (26.3) &    -1.8  & \nodata        & \nodata \\ 
54,701.86..........................  &   0.488  &   105.8 (5.4)  &    13.0  &  -125.6 (6.7)  &    11.6 \\
54,724.65..........................  &   0.447  &   122.3 (3.8)  &    13.0  &  -146.6 (7.9)  &    10.0 \\
54,724.75..........................  &   0.481  &   103.5 (4.4)  &     7.6  &  -137.3 (5.5)  &     3.5 \\
54,725.67..........................  &   0.802  &  -138.8 (3.8)  &    -7.1  &   120.9 (4.7)  &    -6.9 \\
54,725.77..........................  &   0.837  &  -141.6 (4.3)  &     4.9  &   136.6 (5.2)  &    -8.8 \\
54,726.67..........................  &   0.151  &    18.0 (3.9)  &    10.1  &   -60.8 (4.9)  &   -23.8 \\
54,726.77..........................  &   0.185  &    35.9 (3.1)  &    -1.5  &   -83.7 (3.8)  &   -11.9 \\
54,729.68..........................  &   0.201  &    67.0 (3.6)  &    17.0  &   -70.7 (4.5)  &    15.9 \\
54,729.74..........................  &   0.226  &    86.8 (2.7)  &    18.3  &   -89.5 (3.3)  &    19.1 \\
55,040.87..........................  &   0.878  &  -123.0 (6.6)  &    33.2  &   179.4 (8.2)  &    22.6 \\
55,041.67..........................  &   0.155  &    11.9 (28.4) &     0.4  & \nodata        & \nodata \\ 
55,043.67..........................  &   0.856  &  -118.9 (6.8)  &    33.4  &   191.7 (8.3)  &    39.5 \\
55,044.77..........................  &   0.240  &    79.4 (9.0)  &     1.7  &  -107.9 (11.1) &    11.5 \\
55,045.70..........................  &   0.565  &    55.6 (6.4)  &     9.0  &  -117.8 (7.9)  &   -35.2 \\
55,046.68..........................  &   0.906  &  -123.4 (6.1)  &    33.5  &   190.4 (7.5)  &    32.9 \\
55,047.66..........................  &   0.250  &   103.3 (4.4)  &    19.1  &   -90.3 (8.9)  &    36.8 \\

\enddata
\end{deluxetable}

\clearpage

\begin{deluxetable}{
lrrrrrr}
\centering
\rotate
\tabletypesize{\scriptsize}
\tablewidth{0pt}
\tablecaption{Orbital Elements \label{orbparms4.tab}}
\tablehead{
\colhead{Element} &
\colhead{MT311} &
\colhead{MT429} &
\colhead{MT605} &
\colhead{MT696} & 
\colhead{MT720} &
\colhead{MT771}}  
\startdata
$P$ (Days)             & 5.752$\pm$0.005   & 2.9786$\pm$0.0001 & 12.27$\pm$0.01   & 1.4692$\pm$0.0005  & 4.3622$\pm$0.0003    & 2.8635$\pm$0.0002 \\
$e$                    & 0.0 (fixed)       & 0.0 (fixed)       & 0.24$\pm$0.07    & 0.01$\pm$0.02      & 0.35$\pm$0.02	      & 0.05$\pm$0.03 \\
$\omega$ (deg)         & 341$\pm$180       & 339$\pm$180       & 56$\pm$13        & 170$\pm$180        & 297$\pm$3 	      & 41$\pm$29\\
$\gamma$ (\kms)        & 1$\pm$4           & -14$\pm$1         & -12$\pm$2        & -10$\pm$3          & -9$\pm$3 	      & -13$\pm$2 \\
$T_0$ (HJD-2,400,000)  & 52,149 (fixed)    & 51,784.66 (fixed) & 54,653.2$\pm$0.5 & 54729.0$\pm$0.7    & 54,585.80$\pm$0.04   & 54,408.4$\pm$0.2 \\
$K_{1}$ (\kms)         & 88$\pm$5          & $\sim$140         & 44$\pm$3         & 261$\pm$4	       & 173$\pm$5 	      & 139$\pm$5 \\
$K_{2}$ (\kms)         & 118$\pm$9         & $\sim$205         & 77$\pm$5         & 307$\pm$9	       & 242$\pm$7 	      & 164$\pm$6\\
$M_{1}$ (\msun)        & $>$3.0$\pm$0.7\tablenotemark{a} & 7.6  & $>$1.3$\pm$0.3\tablenotemark{a} & $>$15.1$\pm$0.7\tablenotemark{a}	& $>$15.5$\pm$0.8\tablenotemark{a} & $>$4.5$\pm$0.3\tablenotemark{a} \\ 
$M_{2}$ (\msun)        & $>$2.2$\pm$0.4\tablenotemark{a} & 5.2  & $>$0.7$\pm$0.1\tablenotemark{a} & $>$12.8$\pm$0.5\tablenotemark{a}	& $>$11.1$\pm$0.6\tablenotemark{a} & $>$3.8$\pm$0.2\tablenotemark{a} \\ 
$f(m)_1$ (\msun)       & 0.4$\pm$0.1       & \nodata           & 0.10$\pm$0.03    & 2.7$\pm$0.2         & 1.9$\pm$0.2	 & 0.80$\pm$0.08	  \\
$f(m)_2$ (\msun)       & 1.0$\pm$0.3       & \nodata           & 0.5$\pm$0.1      & 5.2$\pm$0.4	        & 5.3$\pm$0.6    & 1.3$\pm$0.1	  \\
S.~C.$_1$              & B2V               & [B3V]             & B0.5V            & O9.5V               & B0--1V 	 & O7V \\
S.~C.$_2$              & B3V               & [B6V]             & [B2.5:V]         & B1:V	        & B1--2V 	 & O9V \\
$a_{1}$sin~$i$ (\rsun) & 10.0$\pm$0.6      & 8.3               & 10.4$\pm$0.8     & 7.6$\pm$0.1         & 14.0$\pm$0.4   & 7.867$\pm$0.004 \\
$a_{2}$sin~$i$ (\rsun) & 13.40$\pm$0.01    & 12.1              & 18.07$\pm$0.02   & 8.921$\pm$0.004     & 19.57$\pm$0.01 & 9.3$\pm$0.3	  \\
$rms_1$ (\kms)         & 12.5              & \nodata           & 7.2    	  & 21.6	        & 17.9		 & 16.8 \\
$rms_2$ (\kms)         & 23.0              & \nodata           & 8.6    	  & 29.5	  	& 25.4 		 & 23.9 \\  
$i$ ($^\circ$) & $\sim$42\tablenotemark{b} & 89 & $\sim$80\tablenotemark{b} &  $\sim$26\tablenotemark{b} &  $\sim$70\tablenotemark{b} &  $\leq$37\tablenotemark{b}   \\
$T_{1}$ (K)            & \nodata           & 19,100            &  \nodata         &  \nodata            &   \nodata      &  \nodata      \\
$T_{2}$ (K)            & \nodata           & 13,700            &  \nodata         &  \nodata            &   \nodata      &  \nodata      \\
$R_{1}$ (\rsun)        & \nodata           & 4.8               &  \nodata         &  \nodata            &   \nodata      &  \nodata      \\
$R_{2}$ (\rsun)        & \nodata           & 3.6               &  \nodata         &  \nodata            &   \nodata      &  \nodata      \\
\enddata
\tablecomments{Calculated errors are located in parentheses.Square brackets indicate spectral types
that have been determined indirectly from the data (i.e., based on the
computed mass ratio rather than the component's spectrum).}
\tablenotetext{a}{Calculated mass is equal to $M$sin$^3i$.}
\tablenotetext{b}{Estimated based on the theoretical masses of the components' spectral types.}

\end{deluxetable}

\clearpage

\begin{deluxetable}{lcr}
\centering
\tabletypesize{\tiny}
\tabletypesize{\scriptsize}
\tablewidth{0pc}
\tablecaption{Ephemeris for MT429 Spectroscopy \label{OC34}}
\tablehead{
\colhead{} & 
\colhead{} & 
\colhead{$V_r$} \\
\colhead{Date (HJD-2,400,000)} &
\colhead{$\phi$} &
\colhead{(\kms)}}
\tablecolumns{3}
\startdata  
51,805.89..........................  &  0.187  &    0.8 (5.2)  \\
52,146.85..........................  &  0.658  &  -28.0 (4.0)  \\
52,161.67..........................  &  0.634  &  -23.1 (4.5)  \\
52,162.84..........................  &  0.029  &   -5.1 (3.9)  \\
53,339.59..........................  &  0.100  &    3.9 (5.7)  \\
53,340.59..........................  &  0.437  &  -15.5 (4.5)  \\
53,989.78..........................  &  0.391  &   -6.7 (6.6)  \\
54,286.91..........................  &  0.148  &    2.2 (7.4)  \\
54,628.88..........................  &  0.957  &  -15.9 (4.5)  \\
54,630.85..........................  &  0.619  &  -22.8 (6.2)  \\
54,699.77..........................  &  0.758  &  -32.0 (3.6)  \\
54,700.79..........................  &  0.099  &   -3.7 (12.8) \\
54,701.75..........................  &  0.423  &  -12.7 (2.5)  \\
55,430.90..........................  &  0.221  &    4.0 (7.6)  \\
55,497.59..........................  &  0.612  &  -32.1 (6.5)  \\
\enddata 
\tablecomments{The radial velocities listed in the table reflect the radial
motion of the B2V component as diluted by the (apparently constant
velocity) spectrum of the brighter B0V component.}
\end{deluxetable}

\clearpage

\begin{deluxetable}{lcccccl}
\tabletypesize{\scriptsize}
\tablewidth{0pc}
\tablecaption{OB Binaries in Cyg OB2 \label{Binaries4}}
\tablehead{
\colhead{} & 
\colhead{} & 
\colhead{} & 
\colhead{P} &
\colhead{} &
\colhead{} &
\colhead{} \\
\colhead{Star} &
\colhead{Type} &
\colhead{S.C.} &
\colhead{(days)} &
\colhead{e} &
\colhead{q} &
\colhead{Ref.}}
\startdata
MT059      & SB1     & O8V $+$ [B]                        & 4.8527 (0.0002)      & 0.11 (0.04)   & 0.22--0.67\tablenotemark{a}  & 1 \\
MT145      & SB1     & O9III $+$ [mid B]                  & 25.140 (0.008)       & 0.291 (0.009) & 0.26--0.63\tablenotemark{a}  & 2 \\
MT252      & SB2     & B2III $+$ B1V                      & 18--19               & \nodata       & 0.8 (0.2)                    & 1 \\
MT258      & SB1     & O8V $+$ [B]                        & 14.660 (0.002)       & 0.03 (0.05)   & 0.18--0.89\tablenotemark{a}  & 1 \\
MT311      & SB2     & B2V $+$ B3V                        & 5.752 (0.005)        & 0.0 (fixed)   & 0.75 (0.07)                  & 3  \\
MT372      & SB1/EA: & B0V $+$ [B2:V]                     & 2.228 (fixed)        & 0.0 (fixed)   & $\sim$0.6                 	& 2,4 \\
MT421      & SB1:/EA & O9V $+$ [B9V--A0V]                 & 4.161   	         & \nodata       & $\sim$0.16--0.19          	& 5 \\
MT429      & SB1/EA  & B0V $+$ [B3V] $+$ [B6V]          & 2.9786 (0.0001) 	 & 0.0 (fixed)   & $\sim$0.68                   & 3,5,6,7 \\
MT605      & SB2     & B1V  $+$ [B2.5:V]                  & 12.27 (0.01)         & 0.24 (0.07)   & 0.57 (0.05)                  & 3 \\
MT696  & SB2/EW/KE   & O9.5V $+$ B0V                      & 1.4692$\pm$0.0005	 & 0.01 (0.02)   & 0.85 (0.03)                  & 3,8 \\
MT720      & SB2     & B0--1V $+$ B1--2V                  & 4.3622 (0.0003)      & 0.35 (0.02)   & 0.71 (0.06)                  & 1,3 \\
MT771      & SB2     & O7V $+$ O9V                        & 2.8635 (0.0002)      & 0.05 (0.03)   & 0.85 (0.05)	             	& 1,3 \\
Schulte 3  & SB2/EA: & O6IV: $+$ O9III                    & 4.7464 (0.0002)      & 0.070 (0.009) & 0.44 (0.08)               	& 1,9 \\
Schulte 5  & SB2/EB  & O7Ianfp $+$ Ofpe/WN9               & 6.6 (fixed)          & 0.0 (fixed)   & 0.28 (0.02)               	& 10,11,12,13, \\
           &         & ($+$ B0V:)                         &                      &               &                              & 14,15,16 \\
Schulte 8a & SB2     & O5.5I $+$ O6:                      & 21.908 (fixed)       & 0.24 (0.04)   & 0.86 (0.04)               	& 17,18 \\
Schulte 9  & SB2     & O5: $+$ O6--7:                     & 2.355~yr             & 0.708 (0.027) & 0.9 (0.1)                   	& 19,20      \\
Schulte 73 & SB2     & O8III $+$ O8III                    & 17.28 (0.03)         & 0.169 (0.009) & 0.99 (0.02)               	& 2      \\
A36        & SB2/EA  & B0Ib $+$ B0III                     & 4.674 (0.004)        & 0.10 (0.01)   & 0.70 (0.06)               	& 2,21,22   \\
A45        & SB2     & B0.5V $+$ B2--3V:                  & 2.884 (0.001)        & 0.273 (0.002) & 0.46 (0.02)               	& 2,22      \\
B17        & SB2/EB: & O7: $+$ O9:                        & 4.0217 (0.0004)      & 0 (fixed)     & 0.75 (fixed)                	& 23,24     \\
\enddata

\tablecomments{Photometric types EW/KE, EA, and EB stand for eclipsing
system of the W UMa type (ellipsoidal; $P<1$ day), Algol type (near
spherical), and $\beta$ Lyr type (ellipsoidal; $P>1$ day)
respectively. The mass ratio for MT421 is calculated using the O star
masses of \citet{FM05} and interpolated AB masses of
\citet{drilling}. The mass ratio associated with the system composed
of MT429b~$+$~MT429c has been indirectly determined from the
photometric data using PHOEBE. Square brackets indicate spectral types
that have been determined indirectly from the data (i.e., based on the
computed mass ratio rather than the component's spectrum).}

\tablenotetext{a}{The range in mass ratio incorporates both
observational and inclination uncertainties}

\tablerefs{
(1) Paper~II; 
(2) Paper~III;
(3) This work;
(4) \citet{Wozniak04};
(5) \citet{PK98}; 
(6) \citet{saida11}
(7) \citet{Henderson2011}
(8) \citet{Rios04};
(9) \citet[][ in prep]{Kiminki2012}; 
(10) \citet{Wilson48}; 
(11) \citet{Wilson51}; 
(12) \citet{Mics53};
(13) \citet{Wal73}; 
(14) \citet{Contreras97}; 
(15) \citet{Rauw99}; 
(16) \citet{Hall74};
(17) \citet{Romano69}; 
(18) \citet{Debeck04};
(19) \citet{Naze08};
(20) \citet{Naze10};
(21) \citet{NSVSa};
(22) \citet{Hanson03};
(23) \citet{Stroud10};
(24) \citet{NSVSb}}

\end{deluxetable}

\clearpage

\clearpage

\clearpage

\clearpage

\clearpage

\clearpage

\clearpage

\clearpage

\clearpage

\clearpage

\clearpage

\clearpage

\clearpage

\clearpage

\clearpage

\clearpage

\end{document}